\documentclass{elsart}
\usepackage{amsmath,amssymb}
\usepackage{graphicx}
\journal{Annals of Physics}
\newcommand{\re}{\mathop{\mathrm{Re}}}
\begin{document}
\begin{frontmatter}
\title{Description of intermediate- and short-range $NN$ nuclear force
within a covariant effective field theory}
\author[Tuebingen]{Amand Faessler},
\author[Moscow]{V.I. Kukulin\corauthref{cor}}
\corauth[cor]{Corresponding author.}
\ead{kukulin@nucl-th.sinp.msu.ru} \and
\author[Dubna]{M.A. Shikhalev}

\address[Tuebingen]{Institut f\"{u}r Theoretische Physik, Universit\"{a}t T\"{u}bingen,
Auf der Morgenstelle 14, D-72076 T\"{u}bingen, Germany}
\address[Moscow]{Institute of Nuclear Physics, Moscow State University, 119899 Moscow, Russia}
\address[Dubna]{Joint Institute for Nuclear Research, Moscow Region, 141980 Dubna, Russia}

\begin{abstract}
The present paper is aimed to developing an effective field theory in the GeV region to describe
consistently both elastic and inelastic $NN$-scattering in a fully covariant way. In this
development we employ our novel $s$-channel mechanism of $NN$-interaction at intermediate and
short ranges assuming a formation of the six-quark bag dressed with basic meson fields such as
$\pi$, $\sigma$, $\rho$ and $\omega$ in the intermediate state. The peripheral part of the
interaction is governed by $t$-channel one- and two-pion exchanges described fairly well by
chiral perturbation theory. The $\sigma$-meson dressing should be especially important for the
stabilization of the bag due to a high coupling constant of the $\sigma$-meson with the bag and
also a strong interaction between the $\sigma$-field and space symmetric quark core
$|\mathrm{s}^6[6]\rangle$. This stabilization of the intermediate dressed bag (which shifts the
mass of the bag downward to $NN$ threshold) will be crucially important for the description of
$\pi$, $2\pi$, $\rho$ and $\omega$-meson production in $NN$-collisions, and, in particular, for
the correct description of $\pi^0$ production near threshold. So that, we expect that the
results of the present work will lead to a deeper understanding the short- and
intermediate-range nuclear forces and internucleon correlations in nuclei.
\end{abstract}
\end{frontmatter}

\section{Introduction: modern status of the theoretical description for the intermediate- and
short-range $NN$ interaction} More than 50 years ago  Hans Bethe has remarked about the nature
of nucleon-nucleon in his famous review ``What Holds the Nucleus Together?'': ``\ldots {\em In
the past quarter century physicists have devoted a huge amount of experimentation and mental
labor to this problem -- probably more man-hours than have been given to any other scientific
question in the history of mankind\/} \ldots".   And this claim is still valid today. Despite of
the enormous efforts of many labs, groups and personal researchers over the world during the
last 50 years and despite of the great progress in experiments and fundamental theory provided
by developments in QCD, we are still not having in our disposal a consistent and reliable
dynamical theory for the $NN$-interaction at intermediate and especially at short distances,
which can form a solid basis for the explanation of numerous precise experimental data in
meson-production in $pp$, $pd$, \ldots collisions, in few-body physics and in many two-proton
photo-emission nuclear experiments, like $\mathrm{A}(\gamma,pp)$, $\mathrm{A}(e,e'pN)$ etc., at
rather high momentum transfer.

We enumerate here some of basic points where the conventional description of the short-range
$NN$-interaction in terms of meson exchanges (e.g. $\sigma$, $\rho$, $\omega$, $\rho-\pi$,
$\omega-\pi$ etc.) fails or looks to be quite inconsistent. First of all, the value of the
$\omega NN$ coupling constant (which is responsible for the short-range $NN$ repulsion and the
spin-orbit splitting) accepted in all OBE-models turns out to be in the range $g^2_{\omega
NN}/4\pi=13.6-20$ which is much larger than the respective SU(3) value ($\sim 5$) while all
other coupling constants used in OBE-models are in a good agreement with the respective SU(3)
predictions. Another evident inconsistency is that almost all meson-exchange nucleon-nucleon
models use a too high value for the tensor-to-vector $\rho$-meson coupling to the nucleon,
$\kappa_\rho\approx 6-7$. At the same time, an analysis of the data on $\pi N$ scattering yields
a much more modest value, $\kappa_\rho\approx 1-3$ for this parameter (see, e.g. \cite{Afnan}).
Thus the short-ranged central and tensor nucleon-nucleon forces in boson-exchange models are
enhanced ''by hand'' to reproduce the experimental $NN$ phase shifts.

Another fundamental difficulty in the traditional OBE-approach, also tightly related to the
short-range nucleon-nucleon interaction, are the very large values of cut-off parameters
$\Lambda_{\pi NN}$, $\Lambda_{\pi N\Delta}$, \ldots in meson-baryon vertices, which are
responsible for short-range limit of the $NN$-interaction. In fact, the values of $\Lambda_{\pi
NN}$, $\Lambda_{\rho NN}$  etc. accepted in traditional OBE-models ($\sim 1.5-2$ GeV) are at
least three times larger than those derived from any dedicated experiments (e.g. from $\pi N$
scattering), or from all fundamental theories ($\sim 0.5-0.7$ GeV). Moreover, the values of
$\Lambda_{\pi NN}$ or $\Lambda_{\rho NN}$ necessary to fit the charged pion production in
$pp$-collision is strongly different from those values needed to fit the {\em elastic \/}
$pp$-scattering. And those both values are different from respective values which are adopted in
$3N$-force operator to fit the $3N$ binding energy within $2N$- and $3N$-force models driven by
one-boson exchange \cite{ref1,ref2,ref3,ref3a}.

\begin{figure}[t]
\includegraphics[height=7cm, angle=0]{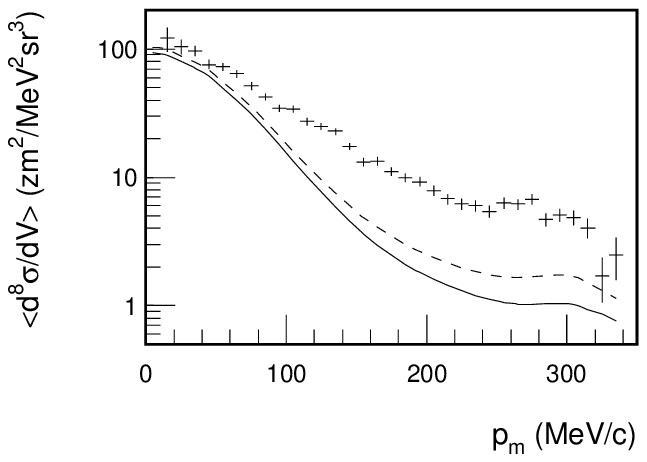}
\caption{Average $\nuc{3}{He}(e,e'pp)$ cross section as a function of the missing momentum $p_m$
for the data measured in kinematic settings LQA and LQV (for the explanation see the text of
Ref. \cite{Groep}) and the (energy, momentum) transfer values ($\omega, q$)=(220 MeV, 305
MeV/{\em c}). The solid and dashed lines represent the results of calculations with a one-body
current operator and including MECs, respectively, using the Bonn-B $NN$ potential.
\label{fig1}}
\includegraphics[width=6.7cm, angle=0]{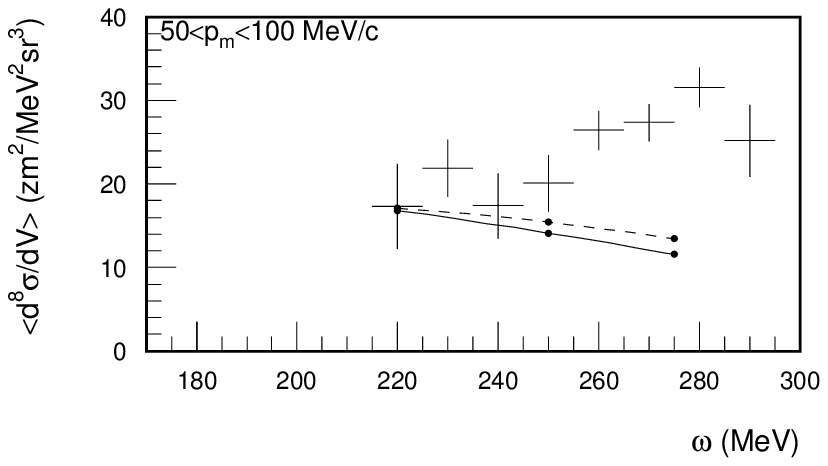}
\includegraphics[width=7cm, angle=0]{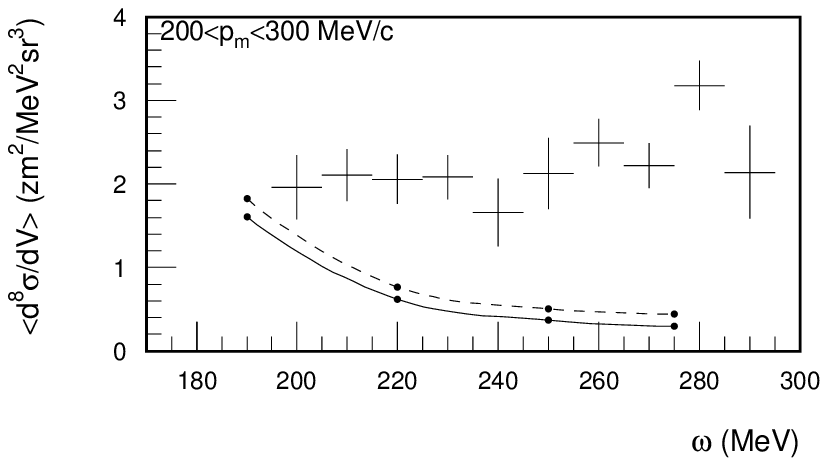}
\caption{Average $\nuc{3}{He}(e,e'pp)$ cross section as a function of the energy
transfer $\omega$ at the value of momentum transfer $q=375$ MeV/{\em c} and $50<p_m<100$
MeV/{\em c} (left) and $200<p_m<300$ MeV/{\em c} (right). The crosses give the experimental data
while the curves represent calculations with a one-body current only (solid), including MECs
(dashed) and including MECs and static isobar currents (dot-dashed).\label{fig1a}}
\end{figure}

A direct consequence of the improper treatment for the short-range nuclear force and respective
short-range two-nucleon correlations in nuclei is a quite serious disagreements between the
predictions of most realistic modern few-body theories and data observed in a few key e.-m.
experiments like $\mathrm{A}(e,e'pp)$, $\mathrm{A}(e,e'pn)$ etc. at rather high momentum and
energy transfers \cite{Ref4}. To illustrate the current situation in this field we present in
Figs. \ref{fig1} and \ref{fig1a} (taken from Ref. \cite{Groep}) the comparison between the most
elaborated modern calculations for $3N$-system (both in bound state and continuum)  which
include also the respective MEC-contributions \cite{Ref13} and the results of recent experiments
at NIKHEF on $\nuc{3}{He}(e,e'pp)n$ reaction.

It is clearly seen from the comparison between theoretical and experimental results that the
underlying mechanism of the reaction should be quite different as compared to that assumed in
the existing theoretical approach (because the energy and momentum dependence derived from the
theory have opposite trends than the experiments showing -- see Fig. \ref{fig1a}). A similar
disagreement has also been observed in electromagnetic knock-out of a $pn$-pair from
$\nuc{3}{He}$ \cite{Ref5} in the Mainz experiments.
 So one can expect a quite different
mechanism for such reactions at rather high momentum transfers as compared to that underlying the
OBE-picture of $2N$ and $3N$ interactions.

The key problem in a consistent theoretical description for such experiments is that the large
incident momentum of a high-energy projectile has to be shared among several nucleons in nuclei,
which requires a
very strong short-range correlation of nucleons \cite{Ref5a}. But the conventional picture of
the correlations induced by the $t$-channel meson exchanges can transfer the high momentum only
with high values of cut-off parameters $\Lambda$, or, alternatively, by incorporation of the
numerous diagrams with multiple rescatterings in intermediate state \cite{Laget}, which are out
of the control.

In the last decade, numerous attempts have been made to describe the intermediate and
short-range $NN$ interaction within the framework of a cluster quark model employing the
Resonating Group Method (RGM) approach \cite{st,VM1,Yaz,Sal}. However all these attempts were
only partially successful and unable to extend the energy range much larger than 300--400 MeV.
Another serious problem with these QCD-inspired approaches is their inability to describe
consistently the inelastic processes, e.g. the one- or two-pion productions. They focused only
on the elastic $NN$ scattering at rather low energies. Thus we have now ``on the market'' two
extreme types of models for the short and intermediate range $NN$ interaction: either the
traditional OBE-type models which do not include any quarks, or the quark-motivated models which
include the mesons only as the carriers of $qq$ interaction \cite{st}, i.e. with no explicit
mesonic degrees of freedom. Thus, with such a limited basis the quark models immediately meet
serious problems in the description of meson production in $NN$ collisions (which do require the
incorporation of mesonic degrees of freedom in {\em an explicit form}).

After the well-known work of Weinberg \cite{Weinberg} the effective field theory (EFT) approach
became very popular and has been applied to the $NN$ interaction and related phenomena in two
forms: (i) in the ``pionless'' form which generalises somehow (in field-theory language) the
effective range expansion \cite{Ref7} and (ii) in the form of Chiral Perturbation Theory (ChPT).
In both directions great progress has been made and full calculations up to high orders are now
feasible. Within the framework of ChPT the calculations have been performed \cite{EntMach,Epel}
(see also \cite{Higa}) for $NN$ elastic scattering with energies up to 290 MeV (in lab. system)
and for deuteron properties. The authors \cite{EntMach} used the ChPT expansion up to fourth
order (i.e. $\mathrm{N}^3\mathrm{LO}$) and have derived the respective $NN$-potential which
leads to the same quality of fit for the $NN$ phase shifts (with $\chi^2\sim 1$) as for the best
modern phenomenological $NN$-potentials. For this fit some 29 free parameters has been employed,
26 of them being related to parametrization of contact terms (in $L\le 2$). These contact terms,
however, are out of the control by chiral theory and thus the description of short-range
interaction is still on a purely phenomenological level.

\begin{figure}[b]
\includegraphics[height=10cm, angle=0]{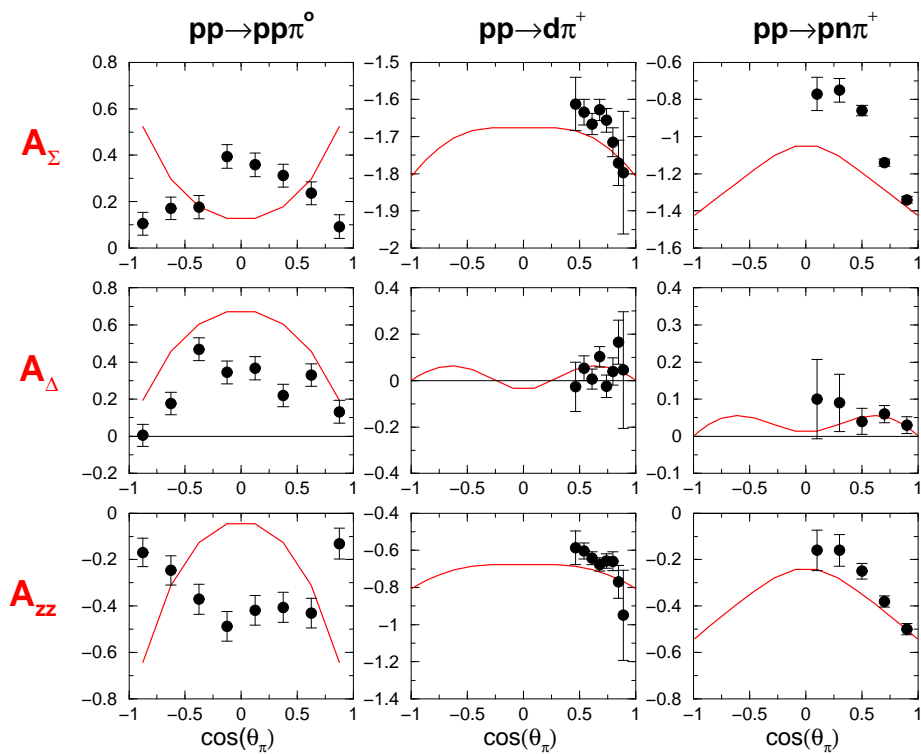}
\caption{Comparison of the model predictions (see the Ref. \cite{Ref9}) to the data
($pp\rightarrow pp\pi^0$) taken from Ref. \cite{Ref8}, ($pp\rightarrow d\pi^+$) Ref.
\cite{Ref9a} and ($pp\rightarrow pn\pi^+$) Ref. \cite{Ref9b}. \label{fig2}}
\end{figure}

\begin{figure}[h]
\begin{center}
\includegraphics[height=60mm,width=65mm,angle=0]{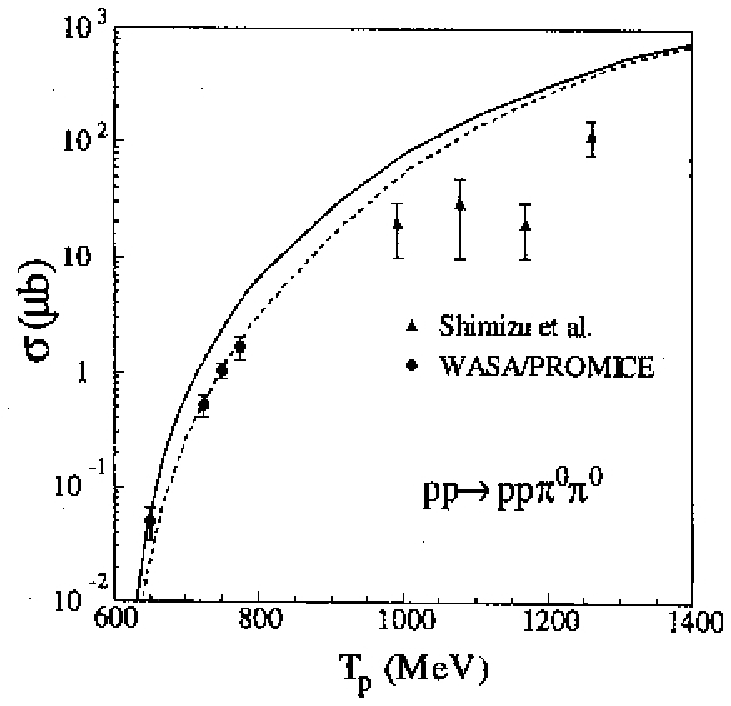}
\includegraphics[height=60mm,width=65mm,angle=0]{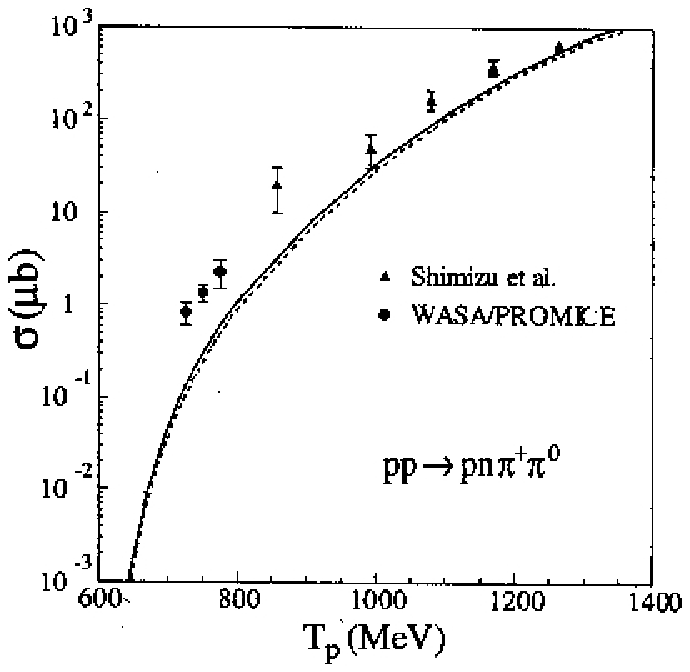}
\end{center}
\caption{The total
cross sections for the two-pion production $pp\rightarrow pp\pi^0\pi^0$ and $pp\rightarrow
pn\pi^0\pi^+$ (quoted from Ref. \cite{Ref10}) shown together with the previously existing data
in the intermediate energy region \cite{Shimizu}. The curves represent theoretical predictions
\cite{Ref10a}. \label{fig3}}
\end{figure}

On the other hand, recent very powerful efforts of the Juelich group to describe quantitatively
the very abundant and precise new data on the neutral pion production cross sections and
analyzing powers in $\pol{p}\pol{p}$-collisions obtained in IUCF \cite{Ref8} have revealed a
rather strong disagreement   between well developed theory based on the meson-exchange approach
and the polarization observables (see Fig. \ref{fig2}) \cite{Ref9}. This means that the
traditional models still fail to describe the $\pi^0$-production in $pp$-collisions (especially
for the spin observables). There are also quite serious problems in the description of {\em
two-pion} production in $pp$-collisions found in recent experiments at Uppsala \cite{Ref10}.
Here the main focus is on the {\em simultaneous description} of $\pi^+\pi^-$, $\pi^0\pi^0$ and
$\pi^+\pi^0$ pair production. While the existing theory is able to predict the data in one
channel, e.g. in the $\pi^+\pi^-$ channel, it fails to give a reasonable description in the
other two channels (see Fig. \ref{fig3}). So, the existing theory seems not to include some
important contribution(s) in the two-pion production sector.

All these problems are tightly interrelated to the short- and intermediate-range $NN$
interaction. So, the understanding of this part of $NN$-interaction should be essentially
improved. It is worth to add here that some interesting new proposal dedicated to the
experimental and theoretical study for short-range $pp$ correlations in $\nuc{3}{He}$ and other
light nuclei have been elaborated very recently by a team in the US \cite{Brod}. But in their
project the authors avoided any reference to microscopic or field-theory approaches and employed
only some phenomenological estimations based on $pp$ scattering in the high-energy region.

On the other hand, the short-range $NN$ correlations are very crucial for our understanding the
nuclear properties at moderate and high momentum transfer and more deep insight into phase
transitions in nuclear matter. It should be stressed here while the existing theory can treat
the averaged nuclear properties like single-particle momentum distribution on the quantitative
level it meets serious problems when explaining the two-body short-range correlations even for
the lightest nuclei like $\nuc{3}{He}$, where we can treat now both bound states and also the
three-nucleon continuum fully consistently (i.e. with complete account of FSI and rescattering
terms). Thus all the above facts (and also other arguments omitted here) and experimental
findings have motivated us to develop some alternative approach for the intermediate- and
short-range nucleon-nucleon force and alternative mechanisms for single- and two-meson
production and also for short-range correlations in nuclei.

The present study is devoted to developing an alternative and new way for the treatment of the
intermediate- and short-range correlations \cite{ref2,Ref6,Ref6a}. This novel approach starts
with an important observation that the $t$-channel nucleon-nucleon interaction induced by
two-pion exchange with full account of $2\pi$ correlation in $I=J=0$ state (i.e. in the $\sigma$
meson channel), being treated consistently, leads inevitably to strong short-range {\em
repulsion} and very weak intermediate-range attraction \cite{Ny,Oset,Kask} instead of the strong
attraction induced by the conventional $\sigma$ meson exchange in OBE-models where the $\sigma$
is treated as a stable particle \cite{Kais}. This amazing result has been obtained recently by
two independent methods \cite{Oset,Kask} and thus looks to be established rather reliably.
Moreover, in the latter work \cite{Kask} it was found that even the $NN$ spin-orbit force
induced by scalar $2\pi$-exchange is almost totally {\em repulsive}, i.e. it is in sharp
contrast to that which is required by the fit to $NN$ phase shifts. So, we need some other
general source for intermediate-range strong attraction and the short-range repulsion for the
central $NN$ force and also  an another source for the strong attractive spin-orbital $NN$
interaction.

The new force model proposed recently by two groups jointly: the Moscow State University group
and the T\"{u}bingen University group \cite{ref2,Ref6,Ref6a,Ref14}, treats the intermediate- and
short-range $NN$-interaction as that proceeding through an intermediate dressed six-quark bag
(i.e. via $s$-channel mechanism):
\[NN\rightarrow (6q+\mathrm{meson\ fields})\rightarrow NN,\]
where the dressing by the meson fields (mainly through the $\sigma$-field) surrounding the dense
six-quark core strongly stabilizes the $6q$-bag and shifts its mass downward to the $NN$
threshold. The physical pattern underlying this new $s$-channel mechanism can be shortly
outlined as follows \cite{Ref14}. Due to a strong attraction of the $\sigma$-meson to the quarks
in the $|\mathrm{s}^6[6]\rangle$ configuration and spherical symmetry of the $\mathrm{s}^6$
six-quark bag, the $\sigma$-field will compress somehow the bag. This bag compression may induce
very short-range scalar diquark correlations in the bag and enhances further the quark-meson
interaction. All these attractive effects will compensate completely the energy losses
originated from $\sigma$-meson generation and rather high kinetic energy of quarks in the bag.
So the net result of this interplay has been estimated to be strongly attractive. Then we
employed the quark-meson microscopic model \cite{Ref6,Ref6a} to estimate these effects and to
construct the respective new model for the $NN$ interaction at short ranges. Our extensive tests
have demonstrated \cite{Ref6a,Ref14,Ref17,Ref18} that this model works very well and makes it
possible to explain many long-standing puzzles in the field, e.g. it uses the OPE terms with
proper cut-off parameters $\Lambda\sim 0.5$ GeV. The new model predicts also the appearance of a
strong three-nucleon scalar force induced by $\sigma$-exchange between the dressed bag and the
third nucleon.

To study the effects of the new $3N$ scalar force when calculating the properties of
$\nuc{3}{H}$ and $\nuc{3}{He}$ nuclei, we have developed \cite{Ref17,Ref18} some multi-component
formalism which made it possible to include both nucleonic and dibaryon components into $3N$
wave functions. It has been found \cite{Ref17} the latter components in $3N$ wave function has a
quite noticeable weight ($\sim 10\% $ or even higher) and these contribute very essentially to
the total $3N$ binding energy. Thus the interplay between two-body and three-body force
contributions peculiar to the conventional force models is changed drastically in the new
approach. However, despite of this, we were still able to describe fully quantitatively, with no
adjustable parameters, the basic static properties of $\nuc{3}{H}$ and $\nuc{3}{He}$ nuclei,
especially the Coulomb displacement energy $E_B({}\nuc{3}{H})-E_B({}\nuc{3}{He})$ which was a
long-standing puzzle in the field. Thus we can summarize here that albeit the new force model
predicts some novel features of nuclear force it leads, at least, for static observables,
largely to the same $3N$ observables as for other modern approaches like CD-Bonn $2N+3N$ UIX or
TM forces. However many other effects in strong or electroweak interaction phenomena will be
described in fully different manner with the new force model.

In previous versions of the model \cite{Ref6a}, we extensively employed the semi-relativistic
quark-meson microscopic model and $\sigma$-meson dressing and have fitted the lower partial $NN$
phase shifts  with the model up to 1 GeV. However, in order to work reliably in the range 1--3
GeV we need a consistent fully relativistic approach.

Another important application of the new force model was a treatment of short-range $N\Delta$
interaction within the same dressed bag model \cite{Obukh}. This part of $N\Delta$ interaction
is at the moment still poorly known, however it can be crucially important in the description of numerous
processes in two- and few-nucleon systems going through $\Delta$-isobar current and accompanied
by a large momentum transfer, like one- or two-pion photoproduction $d(\gamma,\pi^+)$,
$d(e,e'\pi^+\pi^-)$ etc. In our new field theoretical approach the $N\Delta$ and $\Delta\Delta$
channels are considered on the same footing as $NN$ channel by calculating the respective
$N\Delta$ and $\Delta\Delta$ loops of the dressed dibaryon \cite{Ref20}.

On the other hand, the formalism developed makes it possible, in general, to treat the elastic
and inelastic channels in a unified way (see the subsequent section of the article). The only
possible way that meets the above general requirements which the present authors can presently
imagine is a fully covariant field-theoretical approach. Effective field-theory approaches
became very popular in this field in last decade.

Such an effective field approach is ideally suited for our purposes because it includes in an
absolutely natural manner the relativistic formalism and the processes (in intermediate and
final states) with generation of other particles (mesons, isobars etc.). It can deal also quite
naturally with meson and baryon loops in the intermediate states.

Thus we propose in the current work to develop a unified effective field theoretical approach to
treat intermediate- and short-range $NN$-interaction at low and intermediate energies of 1--3
GeV jointly with description of meson production processes at $pp$ and $pn$ collisions. The
clear distinction of the present force model from the existing ones in this aspect is the
appearance of new (dressed) dibaryon components in all nuclei with quite significant weight
$\sim 5-10\% $. This component can absorb, contrary to the conventional single-nucleon picture,
a large momentum and energy. So, many processes associated, e.g. with high momentum transfer,
can proceed easily through this new dibaryon component, the latter being excited by external
probe can produce e.g. a few mesons, or an isobar etc. And these production processes through
the intermediate dressed dibaryon should have much a higher probability than the corresponding
processes by absorption on one nucleon or on a meson-exchange current. Our first calculations
made within the framework of this approach to treat some electromagnetic processes in $2N$ and
$3N$ systems have lead to quite promising  results \cite{Ref17,Ref18,Ref19}.

The structure of the paper is as follows. In Section 2, we explain the new physical model for
intermediate and short-range $NN$ interaction. In Section 3, an effective field-theoretical approach
is developed, which describes in the field-theoretical language the above physical model with
$s$-channel dibaryon generation. Section 4 is devoted to derivation of a relativistic
$NN$-potential using the formalism developed above. The
semi-phenomenological nonrelativistic model proposed by our joint group previously is compared
with the present treatment. In Section 5, we dwell briefly on a generalization of our approach to other hadronic
processes like single- and two-meson production in $pp$-collisions. In Conclusion, we summarize
the content of the work.

\section{Physical model for the intermediate- and short-range $NN$ force}
Here we will outline a multiquark model for the intermediate dressed dibaryon in $s$-channel,
whose formation is assumed to be responsible in our approach for intermediate- and short-range
$NN$ interaction.

We start with an observation done a decade ago \cite{Obukh2} that from two possible space
symmetries, viz. $[3]_x\otimes[3]_x=[6]_x\oplus[42]_x$, the
$|\mathrm{s}^6[6]_x\rangle$-component has comparable projections onto three possible two-cluster
components, viz. $|NN\rangle, |\Delta\Delta\rangle$ and $|CC\rangle$, while the second possible
component $|\mathrm{s}^4\mathrm{p}^2[42]_x\rangle$, which corresponds to the $2\hbar\Omega$
excited six-quark configuration (more exactly a coherent sum of many such components), has a
large projection onto the $|NN\rangle$ channel only. Thus, the fully symmetric part of the total
six-quark wave function $|\mathrm{s}^6[6]_x\rangle$ can be identified with the bag-like
structure while the mixed symmetry part $|\mathrm{s}^4\mathrm{p}^2[42]_x\rangle$ should be
associated with the cluster-like $|NN\rangle$ component. Hence, from the six-quark symmetry
point of view the $2\hbar\Omega$-excited six-quark part corresponds to the proper initial $NN$
channel while the ground $|\mathrm{s}^6[6]_x\rangle$ part includes mainly the intermediate
bag-like components. Therefore the whole $s$-channel process going through the intermediate
dressed dibaryon can be represented by the following chain of steps:
\begin{enumerate}
\item Fusion of two three-quark nucleon clusters into $2\hbar\Omega$ excited six-quark system with the space
symmetry $|\mathrm{s}^4\mathrm{p}^2[42]_x\rangle$\footnote{It can be shown this six-quark
component should be identified with the strongly deformed, i.e. string-like six-quark bag.}.
\item Emission of two $S$-wave pions from the excited mixed symmetry bag with subsequent formation
of the scalar isoscalar $\sigma$-meson due to very strong correlation of these two $S$-wave
pions: $|\mathrm{s}^4\mathrm{p}^2[42]_x\rangle\Rightarrow|\mathrm{s}^6\rangle+(\pi\pi)_{J=I=0}$.
\item Generation of the spherical dense six-quark bag $|\mathrm{s}^6+\sigma\rangle$ in which the
spherical core is surrounded by strong scalar $\sigma$-field that compresses the quark bag due
to strong attraction between $\sigma$-field and quarks.
\end{enumerate}

In this new phase, one has two very important features characterizing the six-quark dynamics of
the bag:
\begin{itemize}
\item partial restoration of the chiral symmetry, that leads to strong reduction in both
$\sigma$-meson and constituent quark masses;
\item essential enhancement of diquark correlations in the bag due to the bag
compression\footnote{It should be reminded that the characteristic size of diquarks is rather
small. For example, in the random instanton liquid model it is around of the instanton size,
i.e. $r_{\mathrm{dq}}\sim 0.4$ fm \cite{Shuryak}.}. This enhancement should lead to additional
energy gain in the process.
\end{itemize}
These two strong effects lead to a strong effective attraction in $NN$-channel at intermediate
range. And this two-stage complicated mechanism replaces in our approach the simple $t$-channel
$\sigma$-exchange in conventional OBE-models (which is invalid as has been proved by recent
studies \cite{Oset,Kask} -- see the Introduction) and is the origin for intermediate range
$NN$-attraction.

However, this dressed bag state, being singlet in color space, has still a large overlap with
the $NN$-channel, so that this state can decay fast back into the initial $NN$-channel. Hence,
this intermediate dibaryon has a large width ($\sim 300-400$ MeV) and thus cannot be visualized
as a resonance-like bump in $NN$ phase shifts or cross sections. Our previous papers
\cite{ref2,Ref6,Ref6a} have demonstrated that this two-stage mechanism is quite able to explain
with reasonable $\sigma qq$ and $\pi qq$ coupling constants $NN$ phase shifts in low partial
waves up to 1 GeV. So that, quite similar to the nucleon dressing by a pion field, the dibaryon
is dressed mainly by the scalar-isoscalar $\sigma$-field and the physical dibaryon should exist
partially in the bare state and in the dressed state.

Resorting to a diquark language for description of the $NN\rightarrow\mathrm{DBS}$ (the dressed
bag state) transition, one can assume that two unpaired quarks from both nucleons form a diquark
pair (of scalar or axial-vector type) and thus, the intermediate stage in the transition is a
state composed from three diquarks. Then such a three-diquark state gets dressed further with
the $\sigma$-field that squeezes the bag. As a consequence, it can become favourable for
diquarks to condense \cite{Rapp,Alford,Huang} at large energies. At the same time, the chiral
symmetry gets partially restored, i.e. we are faced to a phase transition from the usual
chiral-broken phase to the so-called color superconducting phase. In the present study, we are,
of course, far from the regime of color superconductivity but the quark-quark correlations can
still play a significant role already in low energy physics. In some way diquarks can serve as
useful constituents and building blocks in investigating the properties of multi-quark systems
\cite{Wilczek}. Anyway, the generation of the $\sigma$-field in the symmetric $6q$-bag leads to
a large energy gain and the physical mass of the dressed bag shifts strongly downwards.
Effectively it corresponds to an appearance of a strong attractive force in the $NN$-channel.

Another important issue in the suggested mechanism for $NN\rightarrow\mathrm{DBS}$ transition is
a domination of the excited mixed symmetry $|\mathrm{s}^4\mathrm{p}^2[42], LSI\rangle$ in the
initial stage of the transition, so that the resulting $NN$-wave function has a nodal structure.
That the $2\hbar\Omega$ excited mixed symmetry state dominates in the $NN$-wave function over a
fully symmetric non-excited $|\mathrm{s}^6[6]\rangle$ configuration can be argued from various
points of view. One argument is related to the specific spin-flavor nature of $qq$ interaction
in the Goldstone-boson-exchange model \cite{Stancu}. However, in this model for the $qq$ force,
both configurations lie rather high in energy, so that the effective $NN$-interaction in both
symmetries corresponds to repulsion. The second argument is that the $\sigma$-dressing of the
mixed symmetry bag:
$|\mathrm{s}^4\mathrm{p}^2[42]\rangle\rightarrow|\mathrm{s}^6[6]+\sigma\rangle$ leads to much
stronger energy shift of the bare initial $6q$-configuration than the respective shift of the
symmetric state: $|\mathrm{s}^6[6]\rangle\rightarrow|\mathrm{s}^6[6]+\sigma\rangle$, because the
latter process is in close analogy with the Lamb shift (due to interaction of quarks in the bag
with vacuum fluctuations of the scalar field around the bag) while in the first process the
$\sigma$-field is generated via a real deexcitation of two $\mathrm{p}$-shell quarks and their
transition to the $\mathrm{s}$-orbit.
The next section is devoted to the more formal and strict formulation of this qualitative
pattern in terms of a covariant effective field theory.

\section{Effective field theory with dibaryonic degrees of freedom for description of
intermediate- and short-range $NN$ interaction} 
\begin{figure}[b]
\includegraphics[width=130mm, keepaspectratio]{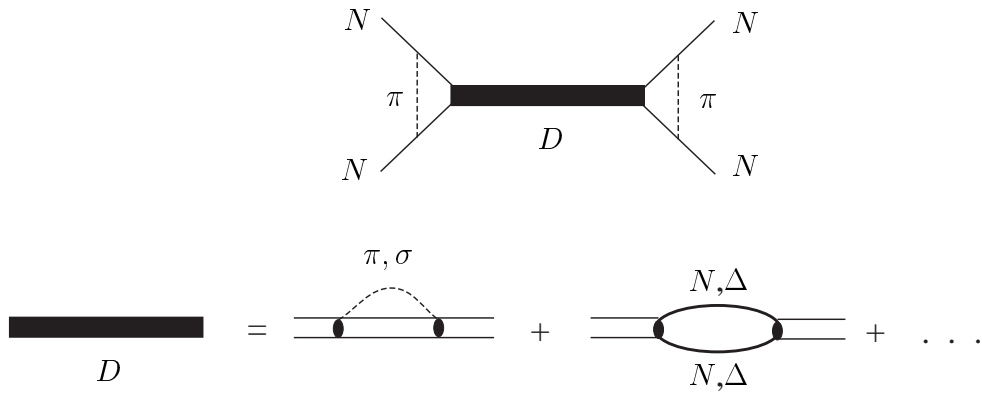}
\caption{The graph in the first row illustrates the $s$-channel
$NN$ interaction through the intermediate dressed dibaryon $D$ (the peripheral one-pion
exchanges are also shown). The graphs in the second row illustrate some possible diagrams for
the dibaryon dressing.\label{fig4}}
\end{figure}

We are aiming to develop a consistent covariant
field-theoretical description for the dibaryon model of nuclear forces, which allows us to go up
to the intermediate energy region 1--3 GeV inaccessible to ChPT. This approach leads to the
important modification of the force model: there arises a consistent scheme for the dressing of
$6q$-bag in terms of effective field theory. Here the additional account of mesonic degrees of
freedom in the DBS leads to an addition of corresponding loops in the complete polarization
operator of the dibaryon, instead of a total replacement of the vertices in the transition
$NN\rightarrow D+m$ ($m=\pi,\sigma,\omega,...$), (as was in our previous version of the model
\cite{Ref6a} and was described briefly in the preceding Section), see Fig. \ref{fig4}.

\begin{figure}[t]
\includegraphics[width=130mm, keepaspectratio]{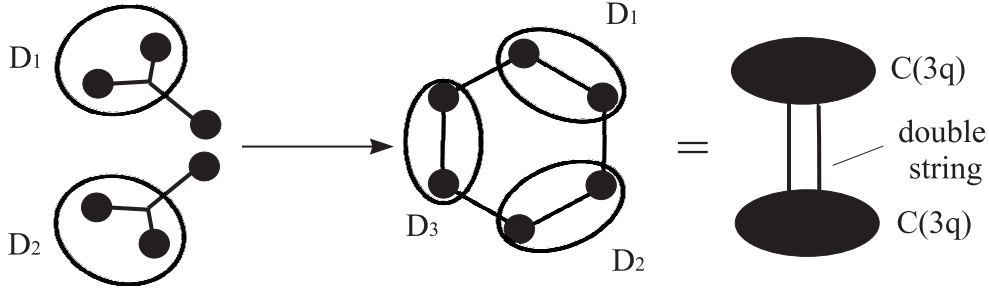}
\caption{Possible mechanism of dibaryon formation. $D_i$
means that there are possible diquark correlations in dibaryon. \label{fig5}}
\end{figure}

It is important to stress that such a field theoretical picture allows to carry out a natural
matching between the field theory approach at low-energies (where ChPT is fully applicable) and
the high energy region. In both regions one can now formulate a field theoretical language. In
such an approach, the peripheral, i.e. the ``external'' part of $NN$-interaction is described by
the conventional $\pi$- and $2\pi$-exchanges in the $t$-channel while the intermediate and short
range interaction is described via generation of an intermediate dressed dibaryon which can be
represented as a color quantum string able to vibrate and rotate. In the present model the
intermediate dibaryon is produced by the color exchange between two quarks belonging to
different nucleons with subsequent formation of a double string (possibly with diquark
clusterization due to a strong $\sigma$-field generated on later stages of the process). The
schematic evolution for this process is depicted in Fig. \ref{fig5}.  To be more specific, we
assume when the two quarks belonging to different $3q$-clusters interact and exchange a color,
this can lead to the formation of either a nonconfined six-quark state (i.e. $NN$, $N\Delta$,
etc.) or a confined $6q$-state locking by a hexagon-type string \cite{Ref20}. The first
non-resonance process is, in fact, analogous in action to conventional meson exchange, while the
second leads to formation of a hidden color $CC$-state. In diquark language, the first mechanism
can be represented by the diagram with two interacting quarks forming a color representation
${\bf 6}$ of SU(3) group, because the one-gluon exchange or instanton induced interaction
between quarks in this channel is repulsive. But if two interacting quarks form the system in a
$\overline{\bf 3}$ color representation, the attractive correlations may lead to patching the
quarks and as a consequence to the diquark formation. In this latter case, the dibaryon can be
modeled as a system of three diquarks connected by Y-type string.

\subsection{General formalism}
Because the value of quark radius of a nucleon is around 0.6 fm and the interaction between
quarks can proceed also at the peripheral regions of both nucleons, the size of the intermediate
dibaryon can be around 1 fm, so that the dibaryon cannot be considered as a point-like object.
Thus, the transition amplitude $NN\rightarrow D(CC)\rightarrow NN$ must be described through the
non-local Lagrangian density:
\begin{equation}
\label{eq0} M_{fi}=\mathrm{i}\langle 4,3|\mathrm{T}\left [e^{\mathrm{i}\int \d x\d
x'\mathcal{L}_{\mathrm{int}}(x,x')}\right ] |2,1\rangle-\mathrm{i}\langle
4,3|\mathrm{I}|2,1\rangle.
\end{equation}
Here T means time-ordering while the bra and ket $|2,1\rangle$ and $\langle 4,3|$ relates to the
initial and final nucleons with 4-momenta $p_2,p_1$ and $p_4,p_3$ respectively. The second term
in (\ref{eq0}), where $\mathrm{I}$ is the unit matrix, corresponds to the propagation of
noninteracting nucleons and thus must be subtracted.

The nonlocal Lagrangian density in Eq. (\ref{eq0}) corresponding to transition from the
two-nucleon to a bare dibaryon state, and the subsequent dibaryon dressing can be written in the
form:
\begin{equation}
\label{eqdd} \mathcal{L}_{\mathrm{int}}(x_1,x_2)=\mathcal{L}_{DN}+\mathcal{L}_{Dm},
\end{equation}
where the Lagrangian {\small\begin{equation} \label{eql3}
\mathcal{L}_{DN}=\widetilde{N}(x_2)\{\Psi(x_1,x_2)V_0(x_1-x_2)+\gamma_5 \gamma^\mu
\Psi_{\mu}(x_1,x_2)V_1(x_1-x_2)\}N(x_1)+ \mathrm{h.c.}
\end{equation}}is odd in dibaryon field operators $\Psi$ and $\Psi_\mu$, and the Lagrangian $\mathcal{L}_{Dm}$
describing the interaction of $6q$-bag with its meson cloud, is even in $\Psi$'s. The function
$\Psi(x_1,x_2)$ describes the dibaryon with spin $S=0$, while $\Psi_\mu(x_1,x_2)$ corresponds to
the $S=1$ dibaryon (in both cases the isospin is assumed to have the value $I=0$, for the case
of isospin $I=1$ we should make a replacement $\Psi\rightarrow\vec{\tau}\vec{\Psi}$ in
(\ref{eqdd})). The bispinor $\widetilde{N}\equiv N^TC^{-1}\gamma_5\mathrm{i}\tau_2$ has the same
transformation properties as the Dirac-conjugated nucleon-field operator
$\overline{N}=N^\dag\gamma_0$ in the Lorentz and isospin groups, $C=\mathrm{i}\gamma_2\gamma_0$
is the charge-conjugation operator, $\tau_2$ is a Pauli matrix, and $N^T$ stands for a
transposed matrix. We use the notation and definitions borrowed from Ref. \cite{Bjorken}.

\begin{figure}[b]
\begin{center}
\includegraphics[width=80mm,keepaspectratio]{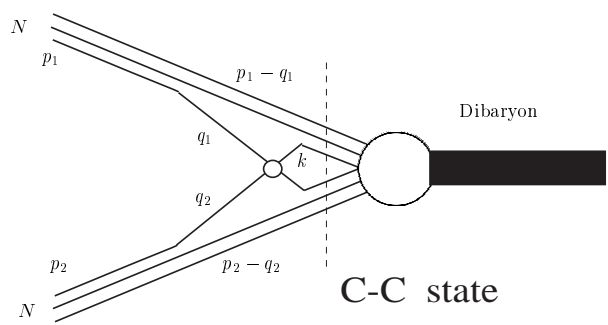}
\end{center}
 \caption{The
vertex for two-nucleon-to-dibaryon transition when two quark, belonging to different nucleons,
interact with each other with subsequent formation of hidden color $CC$-state (or dibaryon)
bound by double string.\label{fig6}}
\end{figure}

The interaction $V(x_1-x_2)$ related to the transition vertex $NN\rightarrow D$ can be
illustrated by the graph shown in Fig. \ref{fig6}. This interaction can be found, in general,
from the quark-microscopic model, e.g. assuming a diquark substructure of the dibaryon. But,
since in this work, we are interesting in developing a general effective field-theory approach
to the dynamics of the intermediate dressed dibaryon we will ignore any details of quark
substructure of the dibaryon. So that, the potential $V$ is taken from the overlap of the inner
nucleonic wave functions (we postpone more elaborated calculation of this transition to the near
future):
\begin{equation}
V(x)=g\int\d ^4yf_{N}(y+x/2;p_1)f_N(y-x/2;p_2),
\end{equation}
where we choose for the $f_N$ the covariant harmonic oscillator wavefunction:
\begin{equation}f_N(x;p)=\frac{\alpha}{\pi}\exp{\left[\frac{\alpha}{2}(x^2-\frac{2(xp)^2}{m_N^2})\right]},\end{equation}
$g$ is an effective coupling parameter and $m_N$ denotes the nucleon mass. In non-relativistic
limit
\begin{equation}f_N(x,p)\approx\frac{\alpha}{\pi}\exp{[-\alpha (x_0^2+{\bf x}^2)/2]},\end{equation}
so we get a covariant form for $V(x)$:
\begin{equation}
\label{potent} V(x)=g\exp{\left[\frac{\alpha}{4}(x^2-\frac{2(xP)^2}{m_N^2})\right]},
\end{equation}
where $P=p_1+p_2$ is the c.m. 4-momentum of two nucleons.

From Eqs. (\ref{eq0}) and (\ref{eql3}), it is evident that the transition amplitude must be even
in dibaryon field operator and at least of the second order in interaction with a nucleon
current. As a result, one has the following expression for $M_{fi}$ (in $I=0$ channel):
\begin{multline}
\label{eq1} M_{fi}=\frac{1}{2!}\sum_{S}\int \d ^4x_1\d ^4x_2\d ^4x_3\d ^4x_4 \langle
4,3|\overline{N}(x_4)\Gamma_S^\dag\overline{\widetilde{N}}(x_3)\times \\ \times
\mathcal{G}^{S}(x_3,x_4;x_1,x_2)\widetilde{N}(x_1)\Gamma_S N(x_2)|2,1\rangle
V_S(x_1-x_2)V_S(x_3-x_4),
\end{multline}
where $\mathcal{G}^S(x_3,x_4;x_1,x_2)$ is an exact dibaryon propagator that takes into account
the dressing of the dibaryon with meson fields. The vertices $\Gamma_S$ are equal:
\begin{equation}\Gamma_S=\Biggl\{ \begin{array}{c}
1,\ \ \ \ \ \ S=0,\\
\gamma_5\gamma_\mu, \ \ S=1.
\end{array}\end{equation}

\begin{figure}[t]
\includegraphics[width=130mm,keepaspectratio]{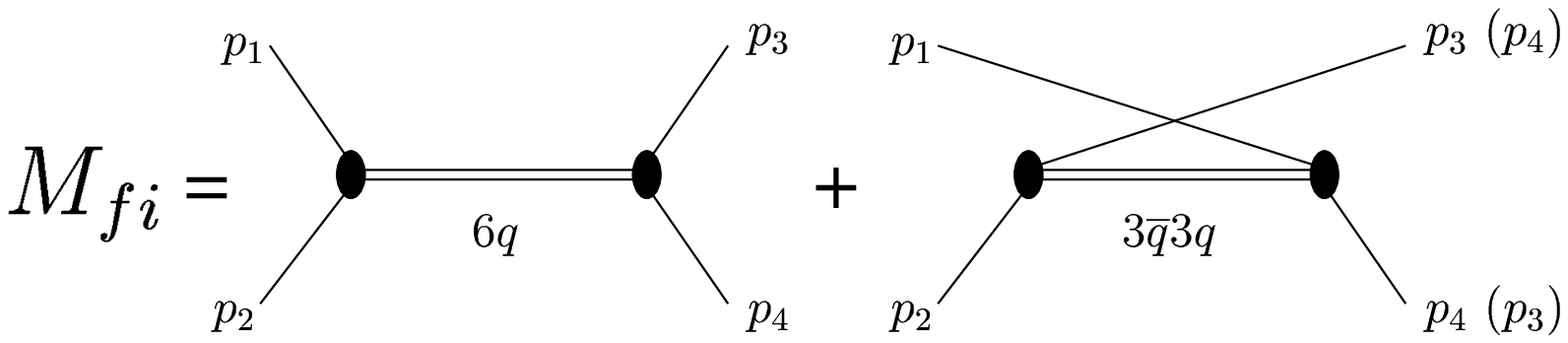}
 \caption{Diagrams representing the amplitude in Eq.
(\ref{eq1}). The first diagram corresponds to $s$-channel dibaryon exchange, while the second
diagram describes the $u$- or $t$-channel exchange of a heavy $\overline{N}N$ meson having an
exotic $3\overline{q}3q$ structure.\label{fig7}}
\end{figure}

The dressed dibaryon propagator is found by solving the respective Dyson equation
\begin{multline}
\label{eq13} \mathcal{G}^{S}(x_3,x_4;x_1,x_2)=G^{S}(x_3,x_4;x_1,x_2)+\\
\int \d ^4x_1'\d ^4x_2'\d ^4x_3'\d ^4x_4'G^{S}(x_3,x_4;x_3',x_4')
\Pi^{S}(x_3',x_4';x_1',x_2')\mathcal{G}^{S}(x_1',x_2';x_1,x_2),
\end{multline}
where
\begin{equation}G^{S}=-\mathrm{i}\langle 0|\mathrm{T}_{X,X'}[\Psi^{S}(x_3,x_4)\Psi^{S}(x_1,x_2)]|0\rangle\end{equation}
is a bare dibaryon propagator and $\Pi^S$ stands for a polarization operator of the
dibaryon\footnote{As for the dressing of the dibaryon we will consider the meson loops and also
intermediate $N\Delta$ and $\Delta\Delta$ loop diagrams. Thus, all other interactions in the
multi-quark system, like one-gluon exchange, instanton-induced, confinement etc. should be
included into the bare dibaryon propagator.}. Here, the time-ordering is performed with respect
to the coordinates $X$ and $X'$ of the dibaryon center of mass. It is worth to note that the
T-ordering on the coordinates (of the two color clusters) $x_3, x_4$ and $x_1, x_2$ in amplitude
$M_{fi}$ corresponds only to the positive and negative signs for the time-like components of the
relative coordinates $x=x_1-x_2$ and $x'=x_3-x_4$ in the integration over $x_i$ ($i=1..4$),
while the mixed ordering, for example, on the coordinates $x_1, x_3$ and $x_2, x_4$ results in
two types of exchange interaction displayed in Fig. \ref{fig7}. The first graph describes the
(resonance-like) $s$-channel exchange by the dibaryon whereas the second graph describes
non-resonance $t$- and $u$-channel exchanges by an exotic dibaryon (which corresponds to a heavy
$N\overline{N}$-meson). This exotic contribution is omitted in Eq.~(\ref{eq1}) and will be
further neglected.

\subsection{Dibaryon wavefunctions}
We consider the dibaryon wavefunctions as a convolution of orbital and total spin parts:
$\Psi(x_1,x_2)=\Phi(x;P)\otimes S(X;P)$, where the orbital part $\Phi(x;P)$ is taken as a
superposition of covariant harmonic oscillator (CHO) eigenfunctions $f(x;P)$ (see below).
Although there is presently no direct connection between the CHO-formalism and QCD, its
application appears quite successful in describing the baryon spectra and the systematics of
$\overline{q}q$ meson states \cite{Ishida}; in addition, it leads to hadron form factors whose
behavior agrees well with experimental data. As for the microscopic basis of the CHO-functions
one should note, that these eigenfunctions are closely related to the solution of the Dirac
equation with a linear vector potential \cite{Ref21} which corresponds to a string-like
interaction (between two colored clusters). The total spin wavefunctions $S(X;P)$ satisfy to the
standard Klein--Gordon equation:
\begin{equation}
\left(\partial^2/\partial X^2_\mu+M_{N_e}^2\right)S(X;P)=0,
\end{equation}
while the $f(x;P)$ is a tensor of rank $N_e$ ($N_e$ is the number of excitation quanta of the
string) in the space $O(3,1)_{\mathrm{Lorentz}}$ and satisfies to the equation:
\begin{equation}
\label{eq6} \mathcal{M}^2(x_\mu,\partial/\partial x_\mu)f(x;P)=M_{N_e}^2f(x;P).
\end{equation}
with the mass operator squared is defined as
\begin{equation}
\label{eq7} \mathcal{M}^2\equiv d\left(\frac{1}{2\mu}\frac{\partial^2}{\partial
x^2_\mu}+U_{\mathrm{conf}}+U_{\mathrm{pert.QCD}}\right),
\end{equation}
here $d=2(m_1+m_2)$, $\mu=m_1m_2/(m_1+m_2)$ and $m_i$ are kinematical masses of  $3q(C)$
clusters on the ends of the string. $U_{\mathrm{conf}}$ is defined by the string potential
energy and also by non-perturbative interactions of instanton type:
\begin{equation}
\label{eq9} U_{\mathrm{conf}}(x)=-(1/2)Kx_\mu^2+C,
\end{equation}
while the perturbative term $U_{\mathrm{pert.QCD}}$ (the effect of which can be reduced mainly
to some deviation of the respective Regge-trajectory from the straight line) is omitted here.
This leads to the following CHO eigenfunctions:
\begin{align}
&f_0(x;P)\equiv\langle
x|G(P)\rangle=\frac{\beta}{\pi}\exp\left[\frac{\beta}{2}
\left(x^2-2\frac{(Px)^2}{P^2}\right)\right],\label{eq9a}\\
&f_{N_e}(x;P)=\langle x|a^\dag_{\nu_{N_e}}\ldots
a^\dag_{\nu_1}|G(P)\rangle, \\
&M_{N_e}^2=dC+({N_e}+1)\Omega=M^2_0+{N_e}\Omega,
\end{align}
where $a_\nu^\dag=(\beta x_\nu+\partial/\partial x_\nu)/(2\beta)^{1/2}$ is the creation operator
for oscillator quanta, $\beta^{-1}=(\mu K)^{-1/2}$ -- squared oscillator radius and
$\Omega=d(K/\mu)^{1/2}$ is the oscillator frequency. For the mass spectrum, we thus obtain a
linear Regge trajectory. The operators $a_\nu^\dag$ are defined in such a way that a physical
state satisfies the condition,
\begin{equation}
\label{cond} P^\nu a_\nu^\dag|\mathrm{phys}\rangle=0,
\end{equation}
which rules out the appearance of an unphysical spectrum with respect to the time variable
\cite{Takabayasi}.

For example, the function corresponding to a single-quantum excitation of the oscillator can be
represented in the form
\begin{equation}f_\mu(x;P)=\sqrt{2\beta}\left(x_\mu-\frac{(Px)P_\mu}{P^2}\right)f_0(x;P).\end{equation}
This function is normalized by the condition
\begin{equation}N_{\mu\nu}\equiv\int f_\mu(x;P)f_\nu(x;P)\d ^4x=-g_{\mu\nu}+\frac{P_\mu P_\nu}{P^2}.\end{equation}
In the case of a two-quantum excitation, two excited states are possible,
\begin{gather}
f_{\mu\nu}(x;P)=\left[g_{\mu\nu}-\frac{P_\mu
P_\nu}{P^2}+2\beta\left(x_\mu-\frac{(Px)P_\mu}{P^2}\right)\left(x_\nu-\frac{(Px)P_\nu}{P^2}\right)\right]f_0(x;P),\\
f_{\mu\mu}(x;P)=\left(3+2\beta x^2-2\beta\frac{(Px)^2}{P^2}\right)f_0(x;P)\nonumber,
\end{gather}
which are normalized as follows:
\begin{multline}
N_{\mu\nu,\alpha\beta}\equiv\int f_{\mu\nu}(x;P)f_{\alpha\beta}(x;P)\d ^4x=\\
\left(g_{\mu\alpha}-\frac{P_\mu P_\alpha}{P^2}\right)\left(g_{\nu\beta}-\frac{P_\nu
P_\beta}{P^2}\right)+\left(g_{\mu\beta}-\frac{P_\mu
P_\beta}{P^2}\right)\left(g_{\nu\alpha}-\frac{P_\nu P_\alpha}{P^2}\right).
\end{multline}

Hereinafter, we shall restrict our consideration only to even-parity dibaryons, including the
states with zero and two quanta only. Moreover, in the current work we shall consider only those
partial waves whose coupling to $N\Delta$-channel is absent or suppressed below 1 GeV in the
lab. system (i.e. we shall limit ourselves to consider in this paper $^1S_0$, $^3S_1-^3D_1$ and
$^3D_2$ partial channels only). Then for corresponding dibaryon wavefunctions
$\Psi^{JN_e}_{LS}(x_1,x_2)$ (where $J$ and $L$ are the total angular and orbital momenta of the
dibaryon), one has: \\ for the $^1S_0$-partial wave:
\begin{equation}\Psi^{00}_{00}(x_1,x_2)=\Phi_{(0\mathrm{s})}(x)S(X),\ \ \ \
\Psi^{02}_{00}(x_1,x_2)=\Phi_{(2\mathrm{s})}(x)S(X),\end{equation} where
$\Phi_{(0\mathrm{s})}(x)=f_0(x)$, $\Phi_{(2\mathrm{s})}(x)=f_{\mu\mu}(x)/\sqrt{6}$ and the
normalization is chosen for a single particle state in a unit volume; \\ for the $^3S_1$-partial
wave\footnote{The tensor coupling of $^3S_1$ and $^3D_1$ partial waves will be considered below
when the dressing procedure will be described.}:
\begin{equation}\Psi^{10}_{01,\mu}(x_1,x_2)=\Phi_{(0\mathrm{s})}(x)S_\mu(X),\ \ \ \
\Psi^{12}_{01,\mu}(x_1,x_2)=\Phi_{(2\mathrm{s})}(x)S_\mu(X);\end{equation} \\ for the
$^3D_1$-partial wave:
\begin{equation}\Psi^{12}_{21,\mu}(x_1,x_2)=\frac{3}{\sqrt{2}}\Phi_{(2\mathrm{d}),\mu\nu}(x)S^\nu(X),\end{equation} where
$\Phi_{(2\mathrm{d}),\mu\nu}(x)=(f_{\mu\nu}(x)-\mathcal{P}^1_{\mu\nu}f_{\alpha\alpha}(x)/3)/\sqrt{15},$
and $\mathcal{P}^1_{\mu\nu}=g_{\mu\nu}-P_\mu P_\nu/P^2$; \\ and for the $^3D_2$-partial wave:
\begin{equation}\Psi^{22}_{21,\mu}(x_1,x_2)=\sqrt{\frac{5}{P^2}}P_\tau\varepsilon^{\mu\beta\rho\tau}\Phi_{(2\mathrm{d}),\alpha\beta}(x)S^\alpha_\rho(X).\end{equation}

Now we can expand the bare dibaryon Green functions into the string eigenstates:
\begin{equation}
G^S(x_3,x_4;x_1,x_2)=-\mathrm{i}\sum_{J,L,N_e}\langle
0|\mathrm{T}_{X,X'}[\Psi^{JN_e}_{LS}(x_3,x_4) \Psi^{JN_e}_{LS}(x_1,x_2)]|0\rangle.
\end{equation}

\subsection{Dressed dibaryon propagators}
In this subsection we will show that, instead of a complicated integral Dyson equation, we can get
its matrix analog (with a very low matrix dimension) for the projection of all propagators and
polarization operators onto the above CHO basis. Simple estimates give the energy value of the
string excitation quantum $\Omega\approx 300-350$ MeV so that it is sufficient to take into
account only two quanta excitations of the dibaryon string when the incident energy in
$NN$-channel does not exceed 1 GeV (in lab. system).

\subsubsection{$^1S_0$ partial channel}
By restricting ourselves to zero and two-quanta excitations of the string, one can write the
bare partial propagator for the $^1S_0$-channel in the form:
\begin{equation}G(x',x;P)=\frac{\Phi_{(0\mathrm{s})}(x')\Phi_{(0\mathrm{s})}(x)}{P^2-M_0^2+\mathrm{i}0}+\frac{\Phi_{(2\mathrm{s})}(x')
\Phi_{(2\mathrm{s})}(x)}{P^2-M_2^2+\mathrm{i}0}+\ldots\end{equation}

Then, by substituting this bare propagator into the Dyson equation (\ref{eq13}) one can get the
following representation for the dressed propagator:
\begin{multline}
\label{eq16}
\mathcal{G}(x',x;P)=\Phi_{(0\mathrm{s})}(x')\Phi_{(0\mathrm{s})}(x)\mathcal{G}_1(P^2)+\Phi_{(0\mathrm{s})}(x')\Phi_{(2\mathrm{s})}(x)\mathcal{G}_2(P^2)+\\
+
\Phi_{(2\mathrm{s})}(x')\Phi_{(0\mathrm{s})}(x)\mathcal{G}_3(P^2)+\Phi_{(2\mathrm{s})}(x')\Phi_{(2\mathrm{s})}(x)\mathcal{G}_4(P^2),
\end{multline}
where
\begin{align}
\stepcounter{equation}
\mathcal{G}_1(P^2)=\frac{P^2-M^2_2-\widetilde{\Pi}_{22}}{(P^2-M^2_0-\widetilde{\Pi}_{11})
(P^2-M^2_2-\widetilde{\Pi}_{22})-\widetilde{\Pi}^2_{12}},\tag{\theequation a}\\
\mathcal{G}_2(P^2)=\mathcal{G
}_3(P^2)=\frac{\widetilde{\Pi}_{12}}{(P^2-M^2_0-\widetilde{\Pi}_{11})
(P^2-M^2_2-\widetilde{\Pi}_{22})-\widetilde{\Pi}^2_{12}},\tag{\theequation b}\\
\mathcal{G}_4(P^2)=\frac{P^2-M^2_0-\widetilde{\Pi}_{11}}{(P^2-M^2_0-\widetilde{\Pi}_{11})
(P^2-M^2_2-\widetilde{\Pi}_{22})-\widetilde{\Pi}^2_{12}},\tag{\theequation c}
\end{align}
in which $\widetilde{\Pi}_{ij}(P^2)$ are polarization operators projected onto the CHO basis:
\begin{align}
\stepcounter{equation} \widetilde{\Pi}_{11}(P^2)=\int \d ^4\xi
\d ^4\zeta\Phi_{(0\mathrm{s})}(\xi;P)\Pi(\xi,\zeta;P)\Phi_{(0\mathrm{s})}(\zeta;P),\tag{\theequation a}\\
\widetilde{\Pi}_{22}(P^2)=\int \d ^4\xi
\d ^4\zeta\Phi_{(2\mathrm{s})}(\xi;P)\Pi(\xi,\zeta;P)\Phi_{(2\mathrm{s})}(\zeta;P),\tag{\theequation b}\\
\widetilde{\Pi}_{12}(P^2)=\int \d ^4\xi \d
^4\zeta\Phi_{(0\mathrm{s})}(\xi;P)\Pi(\xi,\zeta;P)\Phi_{(2\mathrm{s})}(\zeta;P).\tag{\theequation
c}
\end{align}

Then, in order to pursue the renormalization procedure, the dressed propagator
$\mathcal{G}(x',x;P)$ must be diagonalized in the $0\hbar\Omega+2\hbar\Omega$ space. To do this,
one can represent the propagator (\ref{eq16}) in matrix form:
\begin{equation}\mathcal{G}(x',x;P)=\left(\Phi_{(0\mathrm{s})}(x')\ \Phi_{(2\mathrm{s})}(x')\right)OO^T\left(
\begin{array}{cc}
\mathcal{G}_1 & \mathcal{G}_2 \\
\mathcal{G}_3 & \mathcal{G}_4
\end{array}
\right)OO^T\left(
\begin{array}{c}
\Phi_{(0\mathrm{s})}(x)\\ \Phi_{(2\mathrm{s})}(x)
\end{array}\right),\end{equation}
where we introduced the rotation matrix
\begin{equation}O=\left(
\begin{array}{cc}
\cos\chi & \sin\chi \\
-\sin\chi & \cos\chi
\end{array}\right),\ \ \ OO^T=O^TO=1,\ \  \mathrm{det}O=1,\end{equation}
which diagonalizes the matrix propagator. The mixing angle $\chi$, which mixes the different
six-quark configurations $|\mathrm{s}^6,L=0\rangle$ and $|\mathrm{s}^4\mathrm{p}^2,L=0\rangle$,
can be found from the condition
\begin{equation}
\left(\begin{array}{cc} \widetilde{\mathcal{G}}_0 & 0\\ 0 & \widetilde{\mathcal{G}}_2
\end{array}\right)=O^T\left(
\begin{array}{cc}
\mathcal{G}_1 & \mathcal{G}_2\\ \mathcal{G}_3 & \mathcal{G}_4
\end{array}
\right)O,
\end{equation}
Finally, we get for the $\chi(P)$:
\begin{equation}
\label{chi} \chi(P)=\half
\arctan\frac{2\widetilde{\Pi}_{12}(P)}{M^2_2-M^2_0+\widetilde{\Pi}_{22}(P)
-\widetilde{\Pi}_{11}(P)}.
\end{equation}

After the diagonalization of $\mathcal{G}(x',x;P)$, the ``diagonal'' propagators take the
``canonical'' form:
\begin{equation}\widetilde{\mathcal{G}}_{0(2)}(P)=\frac{1}{P^2-M^2_{0(2)(\mathrm{phys})}-\widetilde{\Pi}^
{\mathrm{R}}_{0(2)}(P^2)},\end{equation} where
$\widetilde{\Pi}^{\mathrm{R}}_{0(2)}$ are renormalized matrix elements of polarization operator
\begin{equation}\widetilde{\Pi}^{\mathrm{R}}_{0(2)}(P^2)=\widetilde{\Pi}_{0(2)}(P^2)-(P^2-M^2_{0(2)(\rm{phys})})
\frac{\d \re\widetilde{\Pi}_{0(2)}(P^2)}{\d
P^2}\Biggr|_{P^2=M^2_{0(2)(\mathrm{phys})}},\end{equation} and
\begin{align}\stepcounter{equation}
&\widetilde{\Pi}_0(P)=P^2-M_{0(\mathrm{phys})}^2-\fontsize{15}{\baselineskip}\begin{array}{c}
\frac{(P^2-M_0^2-\widetilde{\Pi}_{11})(P^2-M_2^2-\widetilde{\Pi}_{22})-\widetilde{\Pi}^2_{12}}
{P^2-(M_2^2+\widetilde{\Pi}_{22})\cos^2\chi-(M_0^2+\widetilde{\Pi}_{11})\sin^2\chi-\widetilde{\Pi}_{12}\sin
2\chi},\end{array} \tag{\theequation a}\\
& \widetilde{\Pi}_2(P)=P^2-M_{2(\mathrm{phys})}^2-\fontsize{15}{\baselineskip}\begin{array}{c}
\frac{(P^2-M_0^2-\widetilde{\Pi}_{11})(P^2-M_2^2-\widetilde{\Pi}_{22})-\widetilde{\Pi}^2_{12}}
{P^2-(M_2^2+\widetilde{\Pi}_{22})\sin^2\chi-(M_0^2+\widetilde{\Pi}_{11})\cos^2\chi+\widetilde{\Pi}_{12}\sin
2\chi}.\end{array}\tag{\theequation b}
\end{align}
The physical dibaryon masses are found from the transcendental equations:
\begin{equation}\re\widetilde{\Pi}_0(P^2=M^2_{0(\mathrm{phys})})=
\re\widetilde{\Pi}_2(P^2=M^2_{2(\mathrm{phys})})=0.\end{equation}

\begin{figure}[t]
\includegraphics[width=130mm,keepaspectratio]{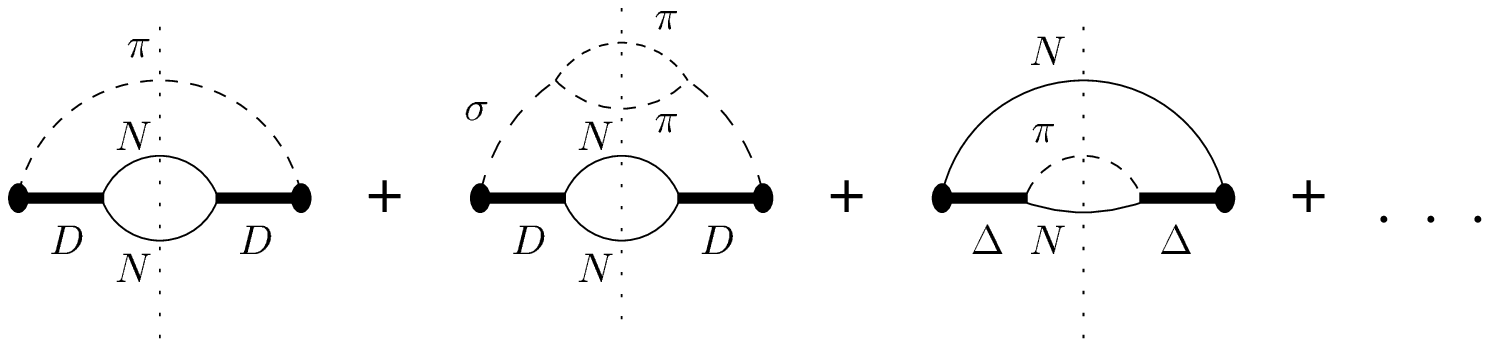}
\caption{Several of the possible loops taken
into account in calculation of polarization operator of dibaryon. They correspond to the dressed
bag state.\label{fig8}}
\end{figure}

We omit here rather lengthy expressions for projected polarization operators and eventual
formulas for the dressed propagators. It is useful, however, to illustrate here the calculations
of polarization operators in different channels by some basic graphs (see Fig. \ref{fig8}).
These graphs include both $\pi$- and $\sigma$-meson dressing of the intermediate dibaryon.

\subsubsection{$^3S_1- {}^3D_1$ mixed channel}
In this case, three six-quark configurations, viz. $|\mathrm{s}^6,L=0\rangle$,
$|\mathrm{s}^4\mathrm{p}^2,L=0\rangle$ and $|\mathrm{s}^4\mathrm{p}^2,L=2\rangle$ can mix with
each other. While the mixing between the first and the second configurations is the same as for
the previous case of $^1S_0$ partial wave, their mixing with the latter $6q$-configuration is
due to the tensor force. Because our focus here is to describe the tensor mixing of $^3S_1$ and
$^3D_1$ channels, induced by the intermediate dibaryon, we neglect now the mixing between the
states with different string excitation energy, which is assumed to be small due to the quite
large energy gap ($2\hbar\Omega\sim 600-700$ MeV) between them.

Thus, the dressed dibaryon propagator can be represented as a sum of two terms:
\begin{equation}
\label{3sd1}
\mathcal{G}_{\mu\nu}(x',x;P)=\mathcal{G}_{\mu\nu}^{(0\hbar\Omega)}(x',x;P)+\mathcal{G}_{\mu\nu}^{(2\hbar\Omega)}(x',x;P),
\end{equation}
where the first and the second terms correspond to $|\mathrm{s}^6\rangle$ and
$|\mathrm{s}^4\mathrm{p}^2\rangle$ dibaryons respectively. Owing to a possible mixing of
$|\mathrm{s}^4\mathrm{p}^2,L=0\rangle$ and $|\mathrm{s}^4\mathrm{p}^2,L=2\rangle$
configurations, the propagator can be represented by a $2\times 2$ matrix. Since we neglect here
the mixing between $|\mathrm{s}^6\rangle$ and $|\mathrm{s}^4\mathrm{p}^2\rangle$ configurations,
for the first term in Eq. (\ref{3sd1}) we have:
\begin{equation}\mathcal{G}^{(0\hbar\Omega)}_{\mu\nu}(x',x;P)=\left(
\begin{array}{cc}
\mathcal{G}^{(N_e=0)}_{\mu\nu} & 0\\
0 & 0
\end{array}\right).\end{equation}
Taking into account the expression for the bare dibaryon propagator of
$|\mathrm{s}^6\rangle$-bag
\begin{equation}G^{(0\hbar\Omega)}_{\mu\nu}(x',x;P)=\left(-g_{\mu\nu}+\frac{P_\mu P_\nu}{M_0^2}\right)
\frac{\Phi_{(0\mathrm{s})}(x';P)\Phi_{(0\mathrm{s})}(x;P)}
{P^2-M_0^2+\mathrm{i}0},\end{equation} one can write the exact propagator, namely the
$\mathcal{G}^{(N_e=0)}_{\mu\nu}$ matrix element, in the form:
\begin{equation}\mathcal{G}^{(N_e=0)}_{\mu\nu}(x',x;P)\equiv\Phi_{(0\mathrm{s})}(x';P)\Phi_{(0\mathrm{s})}(x;P)\widetilde{\mathcal{G}}^{(N_e=0)}_{\mu\nu}(P).\end{equation}
Making further a decomposition $\widetilde{\mathcal{G}}^{(N_e=0)}_{\mu\nu}(P)$ into two
independent tensors of rank 2: $\mathcal{P}^0_{\mu\nu}=P_\mu P_\nu/P^2$ and
$\mathcal{P}^1_{\mu\nu}=g_{\mu\nu}-P_\mu P_\nu/P^2$ that are projectors on the states with spin
0 and 1, and solving the respective matrix Dyson equation, one gets eventually:
\begin{equation}
\label{eqgr}
\widetilde{\mathcal{G}}^{(N_e=0)}_{\mu\nu}(x',x;P)=-\frac{\Phi_{(0\mathrm{s})}(x';P)\Phi_{(0\mathrm{s})}(x;P)}
{P^2-M_{0(\mathrm{phys})}^2-\widetilde{\Pi}^\mathrm{R}(P^2)}\mathcal{P}^1_{\mu\nu}+\frac{\Phi_{(0\mathrm{s})}(x';P)\Phi_{(0\mathrm{s})}(x;P)}{M_0^2-\widetilde{\Pi}_0(P^2)}\mathcal{P}^0_{\mu\nu},
\end{equation}
where in analogy with the $^1S_0$-channel, the renormalized polarization operator is
\begin{equation}
\label{eqgl}
\widetilde{\Pi}^\mathrm{R}(P^2)=M_0^2-M_{0(\mathrm{phys})}^2+\widetilde{\Pi}_1(P^2)-(P^2-M_{0(\mathrm{phys})}^2)
\frac{\d \re\widetilde{\Pi}_1(P^2)}{\d P^2}\Biggr|_{P^2=M^2_{0(\mathrm{phys})}},
\end{equation}
and the physical mass of the dibaryon can be found from the condition
\begin{equation}\re\widetilde{\Pi}_1(P^2=M^2_{0(\mathrm{phys})})=M_{0(\mathrm{phys})}^2-M_0^2.\end{equation}

The expressions for polarization operators projected on CHO basis and spin state  are defined in Eqs.
(\ref{eqgr}) and (\ref{eqgl}) as follows:
\begin{align}\stepcounter{equation}
& \widetilde{\Pi}_1(P^2)=-\frac{1}{3}\int \d ^4\xi
\d ^4\zeta\Phi_{(0\mathrm{s})}(\xi;P)\Pi^{\mu\nu}(\xi,\zeta;P)\Phi_{(0\mathrm{s})}(\zeta;P)\mathcal{P}^1_{\mu\nu},
\tag{\theequation a}\\
& \widetilde{\Pi}_0(P^2)=\int \d ^4\xi \d
^4\zeta\Phi_{(0\mathrm{s})}(\xi;P)\Pi^{\mu\nu}(\xi,\zeta;P)\Phi_{(0\mathrm{s})}(\zeta;P)\mathcal{P}^0_{\mu\nu}.
\tag{\theequation b}
\end{align}

Since the functions $\widetilde{\Pi}_{0(1)}(P^2)$ are analytic, the pole singularity can appear
only in that part of the dibaryon propagator which correspond to spin 1 propagation (that is,
the first term in Eq. (\ref{eqgr})). Because of the dressing of the dibaryon with a meson cloud,
the pole position appears to be shifted from the real axis into the  complex $P^2$ plane and
it corresponds to a resonance state of the dressed six-quark bag. The contribution to the
nucleon-nucleon potential from "unphysical" states of spin zero, which are developed for an
off-shell vector particle and which are inherent in a field-theoretical description of particles
having higher spins, vanishes in the amplitude (\ref{eq1}) due to nucleon current
conservation. In any case, it is of order $1/M_{N_e}^2$ and therefore is very small.

One can write the propagator of the $|s^4p^2\rangle$-bag state in the following matrix form:
\begin{equation}\mathcal{G}_{\mu\nu}^{(2\hbar\Omega)}(x',x;P)=\left(\fontsize{10}{\baselineskip}
\begin{array}{cc}
\Phi_{(2\mathrm{s})}(x')\Phi_{(2\mathrm{s})}(x)\mathcal{G}_{\mu\nu}^{(00)}(P) &
\frac{3}{\sqrt{2}}\Phi_{(2\mathrm{s})}(x')\Phi_{(2\mathrm{d})\beta\nu}(x)\mathcal{G}_{\mu\beta}^{(02)}(P)\\
\frac{3}{\sqrt{2}}\Phi_{(2\mathrm{d})\mu\alpha}(x')\Phi_{(2\mathrm{s})}(x)\mathcal{G}_{\alpha\nu}^{(20)}(P)
&
\frac{9}{2}\Phi_{(2\mathrm{d})\mu\alpha}(x')\Phi_{(2\mathrm{d})\beta\nu}(x)\mathcal{G}_{\alpha\beta}^{(22)}(P)
\end{array}\right).
\end{equation}
Then again, by decomposing the functions
\begin{equation}\mathcal{G}_{\mu\nu}^{(ij)}(P)=-\mathcal{G}_1^{(ij)}(P^2)\mathcal{P}^1_{\mu\nu}+\mathcal{G}_0^{(ij)}
(P^2)\mathcal{P}^0_{\mu\nu}\ \ \ \ \ \ (i,j=0,2),\end{equation} and using condition
(\ref{cond}), we arrive at the following form for the propagator, which has to be diagonalized:
\begin{multline} \label{eq30} \mathcal{G}_{\mu\nu}^{(2\hbar\Omega)}(x',x;P)=
\left(\fontsize{8}{\baselineskip}
\begin{array}{cc}
\Phi_{(2\mathrm{s})}(x')\Phi_{(2\mathrm{s})}(x)\mathcal{G}_0^{(00)}(P^2)\mathcal{P}^0_{\mu\nu} & 0\\
0 & 0
\end{array}\right)-
\left(\fontsize{8}{\baselineskip}
\begin{array}{cc}
\Phi_{(2\mathrm{s})}(x')\mathcal{P}^1_{\mu\alpha} & 0\\
0 & \frac{3}{\sqrt{2}}\Phi_{(2\mathrm{d})\mu\alpha}(x')
\end{array}\right)\times\\
\underbrace{OO}_1\!{}^T\left(\fontsize{8}{\baselineskip}
\begin{array}{cc}
\mathcal{G}_1^{(00)}(P^2) & \mathcal{G}_1^{(02)}(P^2)\\
\mathcal{G}_1^{(20)}(P^2) & \mathcal{G}_1^{(22)}(P^2)
\end{array}\right)
\underbrace{OO}_1\!{}^T\left(\fontsize{8}{\baselineskip}
\begin{array}{cc}
\Phi_{(2\mathrm{s})}(x)\mathcal{P}^1_{\alpha\nu} & 0\\
0 & \frac{3}{\sqrt{2}}\Phi_{(2\mathrm{d})\alpha\nu}(x)
\end{array}\right),
\end{multline}
and
\begin{equation}\left(\begin{array}{cc} \widetilde{\mathcal{G}}_0 & 0\\ 0 & \widetilde{\mathcal{G}}_2
\end{array}\right)=O^T\left(
\begin{array}{cc}
\mathcal{G}_1^{(00)} & \mathcal{G}_1^{(02)}\\ \mathcal{G}_1^{(20)} & \mathcal{G}_1^{(22)}
\end{array}
\right)O.\end{equation}

The first term in Eq. (\ref{eq30}), corresponding to off-shell spin-zero propagation, vanishes
exactly after substituting this expression into the amplitude (\ref{eq1}). By solving the respective
matrix Dyson equation with the bare propagators
\begin{align}
& G^{(2\mathrm{s})}_{\mu\nu}(x',x;P)=-\left(g_{\mu\nu}-\frac{P_\mu
P_\nu}{M_2^2}\right)\frac{\Phi_{(2\mathrm{s})}(x')\Phi_{(2\mathrm{s})}(x)}{P^2-M_2^2+\mathrm{i}0}, \\
& G^{(2\mathrm{d})}_{\mu\nu}(x',x;P)=-\frac{9}{2}\frac{\Phi_{(2\mathrm{d})\mu\alpha}(x')
\Phi_{(2\mathrm{d})\alpha\nu}(x)}{P^2-M_2^2+\mathrm{i}0} ,
\end{align}
we eventually arrive at the following expressions for the functions
$\widetilde{\mathcal{G}}_{0(2)}(P^2)$:
\begin{align}\stepcounter{equation}
& \widetilde{\mathcal{G}}_{0}(P^2)=(P^2-M^2_{(2\mathrm{s})}-\widetilde{\Pi}_0^\mathrm{R}(P^2))^{-1},
\tag{\theequation a}\\
&
\widetilde{\mathcal{G}}_{2}(P^2)=(P^2-M^2_{(2\mathrm{d})}-\widetilde{\Pi}_2^\mathrm{R}(P^2))^{-1},
\tag{\theequation b}
\end{align}
where, as well as just above, after renormalization one has:
\begin{equation}\widetilde{\Pi}^{\mathrm{R}}_{0}(P^2)=\widetilde{\Pi}_{0}(P^2)-(P^2-M^2_{(2\mathrm{s})})
\frac{\d \re\widetilde{\Pi}_{0}(P^2)}{\d P^2}\Biggr|_{P^2=M^2_{(2\mathrm{s})}},\end{equation} (
The same equation is valid for $\widetilde{\Pi}^{\mathrm{R}}_{2}(P^2)$ with a replacement
$M^2_{(2\mathrm{s})}\rightarrow M^2_{(2\mathrm{d})}$ ). Here the diagonalized projected
polarization operators are
\begin{align}\stepcounter{equation}
&
\widetilde{\Pi}_0(P)=P^2-M_{(2\mathrm{s})}^2-\frac{(P^2-M_2^2-\widetilde{\Pi}_{11})(P^2-M_2^2-\widetilde{\Pi}_{22})-\widetilde{\Pi}^2_{12}}
{(P^2-M_2^2)-(\widetilde{\Pi}_{11}\sin^2\chi+\widetilde{\Pi}_{22}\cos^2\chi)-\widetilde{\Pi}_{12}\sin
2\chi}, \tag{\theequation a}\\
&
\widetilde{\Pi}_2(P)=P^2-M_{(2\mathrm{d})}^2-\frac{(P^2-M_0^2-\widetilde{\Pi}_{11})(P^2-M_2^2-\widetilde{\Pi}_{22})-\widetilde{\Pi}^2_{12}}
{(P^2-M_2^2)-(\widetilde{\Pi}_{11}\cos^2\chi+\widetilde{\Pi}_{22}\sin^2\chi)+\widetilde{\Pi}_{12}\sin
2\chi}.\tag{\theequation b}
\end{align}
The mixing angle $\chi$ is defined by the same expression as in the case of the $^1S_0$-channel
(\ref{chi}), where we should make a replacement $M_0\rightarrow M_2$, but with the different
CHO-projected polarization operators:
\begin{align}\stepcounter{equation}
& \widetilde{\Pi}_{11}(P^2)=-\frac{1}{3}\int \d ^4\xi
\d ^4\zeta\Phi_{(2\mathrm{s})}(\xi;P)\Pi^{\mu\nu}(\xi,\zeta;P)\Phi_{(2\mathrm{s})}(\zeta;P)\mathcal{P}^1_{\mu\nu},
\tag{\theequation a}\\
& \widetilde{\Pi}_{12}(P^2)=-\frac{1}{\sqrt{2}}\int \d ^4\xi
\d ^4\zeta\Phi_{(2\mathrm{s})}(\xi;P)\Pi^{\mu\nu}(\xi,\zeta;P)\Phi_{(2\mathrm{d})\mu\nu}(\zeta;P),
\tag{\theequation b}\\
& \widetilde{\Pi}_{22}(P^2)=-\frac{3}{2}\int \d ^4\xi \d
^4\zeta\Phi_{(2\mathrm{d})\mu\alpha}(\xi;P)\Pi^{\alpha\beta}(\xi,\zeta;P)\Phi_{(2\mathrm{d})\beta\mu}(\zeta;P).
\tag{\theequation c}
\end{align}
The physical masses of dibaryons in the oscillator states $(2\mathrm{s})$ and $(2\mathrm{d})$
can be found as usually through the equality
\begin{equation}\re\widetilde{\Pi}_0(P^2=M^2_{(2\mathrm{s})})=
\re\widetilde{\Pi}_2(P^2=M^2_{(2\mathrm{d})})=0.\end{equation}

\subsubsection{$^3D_2$ partial channel}
Now, we shall briefly outline the final formulas for the dressed dibaryon propagator in $^3D_2$
partial channel, as the mostly all steps of the derivation are the same as for the previous
cases.

After substituting the bare $J=2$ dibaryon propagator into the matrix Dyson equation, its
solution can be found in the following form:
\begin{equation}
\label{3d2}
\mathcal{G}_{\mu\nu}(x',x;P)=5\epsilon^{\mu\beta\rho\tau}\epsilon^{\nu\beta_1\rho_1\tau_1}\mathcal{P}^2_{\rho\alpha,\rho_1\alpha_1}\frac{P_\tau
P_{\tau_1}}{P^2}\frac{\Phi_{(2\mathrm{d})\alpha\beta}(x')\Phi_{(2\mathrm{d})\alpha_1\beta_1}(x)}{P^2-M^2_{2(\mathrm{phys})}-\widetilde{\Pi}^\mathrm{R}(P^2)}.
\end{equation}
Here
\begin{equation}\mathcal{P}^2_{\rho\alpha,\rho_1\alpha_1}\equiv\half \mathcal{P}^1_{\rho\rho_1}\mathcal{P}^1_{\alpha\alpha_1}+
\half\mathcal{P}^1_{\rho\alpha_1}\mathcal{P}^1_{\alpha\rho_1}-\frac{1}{3}\mathcal{P}^1_{\rho\alpha}\mathcal{P}^1_{\rho_1\alpha_1},\end{equation}
is a projection operator onto the momentum $J=2$ state. We omit in Eq. (\ref{3d2}) the terms
corresponding to off-shell lower spin propagation of the dibaryon, since they vanish completely
in the amplitude $M_{fi}$ due to nucleon current conservation and condition (\ref{cond}) as
well. We would like to note here, that this is a general result, independent of the value of
total angular momentum $J$. So, our amplitude is free from the "unphysical" lower spin state
contributions inherent in field-theoretical formulations for higher ($J\geq 1$) spin particles.

For the renormalized projected polarization operator, we have
\begin{equation}
\widetilde{\Pi}^\mathrm{R}(P^2)=M_2^2-M_{2(\mathrm{phys})}^2+\widetilde{\Pi}(P^2)-(P^2-M_{2(\mathrm{phys})}^2)
\frac{\d \re\widetilde{\Pi}(P^2)}{\d P^2}\Biggr|_{P^2=M^2_{2(\mathrm{phys})}},
\end{equation}
where \begin{multline}
\widetilde{\Pi}(P^2)=\epsilon^{\mu\beta\rho\tau}\epsilon^{\nu\beta_1\rho_1\tau_1}\mathcal{P}^2_{\rho\alpha,\rho_1\alpha_1}\frac{P_\tau
P_{\tau_1}}{P^2}\times\\
\int \int \d ^4\xi \d
^4\zeta\Phi_{(2\mathrm{d})\alpha\beta}(\xi;P)\Pi^{\mu\nu}(\xi,\zeta;P)\Phi_{(2\mathrm{d})\alpha_1\beta_1}(\zeta;P),
\end{multline}
and the physical $^3D_2$-dibaryon mass is extracted from the condition:
\begin{equation}\re\widetilde{\Pi}(P^2=M^2_{2(\mathrm{phys})})=M_{2(\mathrm{phys})}^2-M_2^2.\end{equation}

\subsubsection{Some remarks on polarization operators}
In principle, the dibaryon polarization operator $\Pi(\xi,\zeta;P)$ (or
$\Pi_{\mu\nu}(\xi,\zeta;P)$) is determined in even partial waves mainly by the $D\sigma$ and
$D\pi$ loops, while its imaginary part is largely determined by the $NN$ loops and these loops
are responsible for dibaryon decay back to two real nucleons. Hence, in our case, if we want to
construct the $NN$-potential, we must exclude coupling to an intermediate nucleon-nucleon state.
This is because of the fact that in the scattering equation (e.g. Bethe--Salpeter or any
quasipotential equation), such channels are automatically taken into account by virtue of the
requirement that the scattering amplitude satisfies two-particle unitarity. Therefore, the main
contribution to the polarization operator comes from the dressing of the dibaryon with its meson
cloud. This interaction induces a transition of the dibaryon (featuring $N_e\geq 2$
string-excitation quanta) to the ground state $|\mathrm{s}^6\rangle$ at low energies. At the
same time, the interaction with $\rho$, $\omega$ and the other mesons adds a much smaller
contribution because of their rather large mass and due to the fact that these mesons can
propagate only in odd partial waves with respect to the dibaryon $|\mathrm{s}^6\rangle$ or in
even partial waves with respect to the $1\hbar\Omega$-excited configuration
$|\mathrm{s}^5\mathrm{p}\rangle$, which has one excitation quantum $\Omega$.

Demanding the parity and total-isospin conservation, one can write the transitions between different
$(S,I)^P$
states that are to be accounted for in polarization operators under consideration:
\begin{align}
& (0,1)^+\longrightarrow(0,1)^+|s^6\rangle+\sigma(L=\mathrm{even}),\nonumber \\
& (0,1)^+\longrightarrow(1,0)^+|s^6\rangle+\pi(L=\mathrm{odd}),\nonumber \\
& (1,0)^+\longrightarrow(1,0)^+|s^6\rangle+\sigma(L=\mathrm{even}), \label{tens}\\
& (1,0)^+\longrightarrow(0,1)^+|s^6\rangle+\pi(L=\mathrm{odd}),\nonumber \\
& (1,0)^+\longrightarrow(2,1)^+|s^6\rangle+\pi(L=\mathrm{odd})\nonumber.
\end{align}
Here, we emphasize an important fact that follows directly from the above quantum numbers for
the allowed transitions: the $s$-channel transitions forming the intermediate states of the
dressed dibaryon are in a full correspondence with ordinary $t$-channel transitions. {\em It is precisely for
this reason that the $s$-channel mechanisms considered here make those contributions to various
high-momentum-transfer processes that were missing in the traditional OBE-models.}

We would also like to indicate that, as it is evident from (\ref{tens}), a $(2,1)^+$ dibaryon
state may be a result of the dressing of the bare dibaryon $(1,0)^+$ with a pion cloud where the
pion is in $P$- (or other odd) partial wave in respect to the six-quark core. Since the
color-dependent interaction between quarks (that produces a hidden $CC$ state) of tensor type is
rather weak, the decay of this state directly to the nucleon-nucleon channel should be strongly
suppressed, but on the other side, the $(2,1)^+$ state has quite a large coupling to the
$N\Delta$-channel. It is well known that just the behavior of the $^1D_2$ phase shift in $NN$
scattering (which corresponds to the $(2,1)^+$ dibaryon state) is indicative of the presence of
a strong $N\Delta$ correlation, which is sizable even at energies of about 400 MeV in the
laboratory frame. In contrast to the conventional meson-exchange models, where the formation of
this state in the $N\Delta$ channel is due to $t$-channel pion exchange and requires hard pion
form factors, our model proposes a totally new mechanism that is responsible for the formation
of a $N\Delta$ state at short and intermediate distances and which is caused by the dressing of
the dibaryon with the pion field; that is, {\small
\begin{equation}(0,1)^+\longrightarrow(1,0)^+|s^6\rangle+\pi(L=\mathrm{odd})\longrightarrow(2,1)^+
\longrightarrow N\Delta(L_{N\Delta}=0,S=2,I=1).\end{equation}} Thus, the assumption of an
intermediate dressed dibaryon gives a very natural explanation for strong coupling between $NN$
and $N\Delta$ channels in some partial waves.

We also note that, in the present model, the polarization operators (and consequently the
$NN$-potential as well) are complex-valued functions. Their imaginary parts are related to open
inelastic channels and are determined by the discontinuities of these quantities at the unitary
cut in the complex energy plane. In the example considered here, we have two inelastic channels
corresponding to real intermediate states, $NN\pi$ and $NN\pi\pi$, where the latter is
associated with the mechanism of $\sigma$-meson decay to two real pions.

\section{Derivation of relativistic $NN$ potential}

Finally we illustrate here how to derive the relativistic $NN$ potential from the formalism
developed. As an example, we consider here the potentials in the lowest partial waves, namely
the isotriplet $^1S_0$  and isosinglet $^3S_1$-$^3D_1$ channels. To this aim we use, in the
amplitude $M_{fi}$ (\ref{eq1}), the relative 4-momenta of nucleons $q=(p_1-p_2)/2$ and
$q'=(p_3-p_4)/2$, and employ of the relation between bispinors with positive and negative
energies
\begin{equation}\label{bispin}
v(q)=C\overline{u}^T(q),\ \ \ \ \ \overline{v}(q)=-u^T(q)C^{-1}.
\end{equation}
Moreover, the transition amplitude taken in the form (\ref{eq1}) is more appropriate for the
reaction $N+\overline{N}$ in the crossed channel. Thus to pass to the $NN$ channel one should
employ the Fierz transformation:
\begin{equation}
[\overline{\psi}\Gamma_i\phi][\overline{\phi}\Gamma_j\psi]=\frac{1}{16}\sum_{k,l=1}^{16}
\mathrm{Tr}(\Gamma_i\Gamma_k\Gamma_j\Gamma_l)[\overline{\psi}\Gamma^k\psi][\overline{\phi}\Gamma^l\phi],
\end{equation}
where $\psi$ and $\phi$ denotes the Dirac bispinors and $\Gamma_i$ corresponds to one of the 16
Dirac matrices $\{I,\gamma_5,\gamma_\mu,\gamma_5\gamma_\mu,\gamma_\mu\gamma_\nu\}$ forming a
basis in the space of $4\times 4$ matrices.

\subsection{$NN$ potential in the $^1S_0$ partial wave}
The $NN$ potential in this channel can now be written in the following form (separating out the
$\delta$-function factor, arising from the integration in Eq. (\ref{eq1}) on the coordinate of
the dibaryon center of mass):
\begin{equation}
\label{eq24} V^{({}^1S_0)}(q',q;P)=-2[\overline{u}(-{\bf q}')\gamma_5v({\bf q}')]
[\overline{v}({\bf q})\gamma_5u(-{\bf q})]F^{(00)}(q',q;P),
\end{equation}
where the factor 2 in the potential arises due to antisymmetrization over nucleon
variables\footnote{We omitted here the isospin factor that is equal 1 for $pn$ scattering and 2
for $pp$ scattering, since in contrast to the $t$-channel interaction, where antisymmetrization
occurs at the level of Feynman diagrams, it is achieved here even at the level of vertices, with
the result that the amplitude $M_{fi}$ appears to be automatically antisymmetrized. In
calculating the $pp$ phase shifts, it must therefore be divided by a factor of 2 (whereupon we
obtain isotopic invariance) with the net result being the same as for the $pn$ potential.}.
$F^{(00)}$ plays the role of the invariant amplitude. We omit here rather lengthy expression for
the $F^{(00)}$ in the case of mixing $0\hbar\Omega+2\hbar\Omega$ states, which is
straightforward to write down employing the results of preceding Section for the dibaryon
propagator. For brevity, we present here the simplified formula for the case of the vanishing
mixing, i.e. we put here $\chi=0$:
\begin{equation}F^{(00)}(q',q;P)=\widetilde{\mathcal{G}}_1(P^2)\widetilde{\Phi}_{(0\mathrm{s})}(q',P)\widetilde{\Phi}_{(0\mathrm{s})}(q,P)+
\widetilde{\mathcal{G}}_2(P^2)\widetilde{\Phi}_{(2\mathrm{s})}(q',P)\widetilde{\Phi}_{(2\mathrm{s})}(q,P),\end{equation}
where
\begin{equation}
\label{formf} \widetilde{\Phi}_{(i)}(q,P)=\int \d ^4x\exp{(\mathrm{i}qx)}V(x)\Phi_{(i)}(x;P).
\end{equation}
Substituting further the expressions for $NN\rightarrow D$ transition potential
(\ref{potent}) and dibaryon wave function into the formula for the dibaryon formfactor
$\widetilde{\Phi}_{(i)}$ one gets:
\begin{equation}\widetilde{\Phi}_{(0\mathrm{s})}(q,P)=g\frac{4\pi\beta}{(\beta+\alpha/2)^2}\exp\left(\frac{q^2-2(qP)^2/P^2}
{2\beta+\alpha}\right),\end{equation}
\begin{equation}\widetilde{\Phi}_{(2\mathrm{s})}(q,P)=\frac{1}{\sqrt{6}}\left[3-\frac{6\beta}{\beta+\alpha/2}-\frac{2\beta}
{(\beta+\alpha/2)^2}(q^2-\frac{(qP)^2}{P^2})\right]\widetilde{\Phi}_{(0\mathrm{s})}(q,P).\end{equation}

Then, using some properties of the Dirac matrices after the Fierz transformation one can write:
\begin{multline}
\label{scalar} [\overline{u}(-{\bf q}')\gamma_5v({\bf q}')] [\overline{v}({\bf
q})\gamma_5u(-{\bf q})]=\\ = -\frac{(1-\vec{\sigma}_1\vec{\sigma}_2)}{2}W \left(1+\frac{{\bf
q}'{}^2}{(E_{q'}+m_N)^2}\right)\left(1+\frac{{\bf q}^2}{(E_{q}+m_N)^2}\right)\!,
\end{multline}
where the factor is $W=(E_{q'}+m_N)(E_q+m_N)/4m_N^2$, $E_q=({\bf q}^2+m_N^2)^{(1/2)}$ and the
spin operator $\vec{S}=(\vec{\sigma}_1+\vec{\sigma}_2)/2$ corresponds to the total spin of the
$NN$ system.

Further, if one employs the partial-wave expansion of the $NN$-potential:
\begin{equation}V({\bf q}',{\bf q};P)=\sum_{JSI\atop LL'}V^{JSI}_{LL'}({\bf q}',{\bf q};P),\end{equation}
with
\begin{equation}V^{JSI}_{LL'}({\bf q}',{\bf q};P)=(4\pi)^2\mathrm{i}^{L-L'}\sum_M\mathcal{Y}^{JM}_{L'S}({\bf n}_{q'})V^{JSI}_{LL'}
(|{\bf q}'|,|{\bf q}|;P)\mathcal{Y}^{JM\dag}_{LS}({\bf n}_{q}),\end{equation} where
\begin{equation}\mathcal{Y}^{JM}_{L'S}({\bf n}_{q})=\sum_{M_LM_S}C^{JM}_{LM_LSM_S}Y_{LM_L}({\bf
n}_{q})|SM_S\rangle\end{equation} represents a standard spin-angular part,
 one finally gets in the channel $J=L=S=0,$ $I=1$ the potential in the form:
{\small\begin{equation} \label{eq21} V^{001}_{00}(|{\bf q}'|,|{\bf
q}|;P)=\frac{W}{\pi}\left(1+\frac{{\bf q}'{}^2}{(E_{q'}+m_N)^2}\right)\left(1+\frac{{\bf
q}^2}{(E_{q}+m_N)^2}\right)F^{(00)}(|{\bf q}'|,|{\bf q}|;P).
\end{equation}}

\subsection{The case of the $^3S_1$-$^3D_1$ mixed channel}
According to the results for the dibaryon propagator in the mixed channel, the $NN$ potential should
be a $2\times 2$ matrix:
\begin{equation}
\label{eq24a} V^{({}^3SD_1)}(q',q;P)=-2[\overline{u}(-{\bf q}')\gamma^\mu v({\bf q}')]
[\overline{v}({\bf q})\gamma^\nu u(-{\bf q})]F_{\mu\nu}(q',q;P),
\end{equation}where the elements of the matrix tensor $F_{\mu\nu}(q',q;P)$ are
\begin{align}\stepcounter{equation}
& F^{(00)}_{\mu\nu}=\Bigl[\widetilde{\mathcal{G}}(P^2)\widetilde{\Phi}_{(0\mathrm{s})}(q',P)
\widetilde{\Phi}_{(0\mathrm{s})}(q,P)+\nonumber \\
&\hspace{3cm}\left(\widetilde{\mathcal{G}}_0(P^2)\cos^2\chi+\widetilde{\mathcal{G}}_2(P^2)\sin^2\chi\right)
\widetilde{\Phi}_{(2\mathrm{s})}(q',P)
\widetilde{\Phi}_{(2\mathrm{s})}(q,P)\Bigr]\mathcal{P}^1_{\mu\nu},\tag{\theequation a}\\
&F^{(02)}_{\mu\nu}=\frac{3}{\sqrt{2}}\widetilde{\Phi}_{(2\mathrm{s})}(q',P)\widetilde{\Phi}_{(2\mathrm{d})\mu\nu}(q,P)
\left[\widetilde{\mathcal{G}}_2(P^2)-\widetilde{\mathcal{G}}_0(P^2)\right]\sin\chi\cos\chi,\tag{\theequation b}\\
&F^{(20)}_{\mu\nu}=\frac{3}{\sqrt{2}}\widetilde{\Phi}_{(2\mathrm{d})\mu\nu}(q',P)\widetilde{\Phi}_{(2\mathrm{s})}(q,P)
\left[\widetilde{\mathcal{G}}_2(P^2)-\widetilde{\mathcal{G}}_0(P^2)\right]\sin\chi\cos\chi,\tag{\theequation c}\\
&F^{(22)}_{\mu\nu}=\frac{9}{2}\widetilde{\Phi}_{(2\mathrm{d})\mu\alpha}(q',P)\widetilde{\Phi}_{(2\mathrm{d})\alpha\nu}(q,P)
\left(\widetilde{\mathcal{G}}_0(P^2)\sin^2\chi+\widetilde{\mathcal{G}}_2(P^2)\cos^2\chi\right),\tag{\theequation
d}
\end{align} where formfactor $\widetilde{\Phi}_{(2\mathrm{d})\mu\nu}(q,P)$ is defined by Eq. (\ref{formf})
and  one has in explicit form:
\begin{equation}
\widetilde{\Phi}_{(2\mathrm{d})\mu\nu}(q,P)=\frac{2\beta}{\sqrt{15}(\beta+\alpha/2)^2}q_\alpha
q_\tau\left(\frac{1}{3}\mathcal{P}^1_{\mu\nu}\mathcal{P}^1_{\alpha\tau}-\mathcal{P}^1_{\mu\alpha}\mathcal{P}^1_{\tau\nu}\right)\widetilde{\Phi}_{(0\mathrm{s})}(q,P).
\end{equation}

After making the respective Fierz transformations, one arrives eventually at the following
expressions for the matrix elements $V^{(ij)}$: \begin{align} \label{eq53}\stepcounter{equation}
V^{(00)}(q',q,P)=(3+\vec{\sigma}_1\vec{\sigma}_2)W&\fontsize{11}{\baselineskip}\begin{array}{c}\left(1+\frac{{\bf
q}'{}^2+S_{12}({\bf q }')}{3(E_{q'}+m_N)^2}\right)\left(1+\frac{{\bf q}^2+S_{12}({\bf
q})}{3(E_{q}+m_N)^2}\right)\end{array}F^{(00)}(q',q;P),\tag{\theequation
  a}\nonumber\\
V^{(02)}(q',q,P)=(3+\vec{\sigma}_1\vec{\sigma}_2)W&\left(1+\frac{{\bf q}'{}^2+S_{12}({\bf q
}')}{3(E_{q'}+m_N)^2}\right)\times\nonumber \\ &\left(1+\frac{{\bf q}^2+S_{12}({\bf
q})}{3(E_{q}+m_N)^2}\right)\frac{S_{12}({\bf
q})}{6{\bf q}^2}F^{(02)}(q',q;P),\tag{\theequation b}\nonumber
\end{align}
\begin{align}
V^{(20)}(q',q,P)=(3+\vec{\sigma}_1\vec{\sigma}_2)W&\frac{S_{12}({\bf q}')}{6{\bf
q}'{}^2}\left(1+\frac{{\bf q}'{}^2+S_{12}({\bf q }')}{3(E_{q'}+m_N)^2}\right)\times\nonumber \\
&\phantom{\frac{S_{12}({\bf q}')}{6{\bf q}'{}^2}}\left(1+\frac{{\bf q}^2+S_{12}({\bf
q})}{3(E_{q}+m_N)^2}\right)F^{(20)}(q',q;P),\tag{\theequation c}\\
V^{(22)}(q',q,P)=(3+\vec{\sigma}_1\vec{\sigma}_2)W&\frac{S_{12}({\bf q}')}{6{\bf
q}'{}^2}\left(1+\frac{{\bf q}'{}^2+S_{12}({\bf q }')}{3(E_{q'}+m_N)^2}\right)\times\nonumber  \\
&\left(1+\frac{{\bf q}^2+S_{12}({\bf q})}{3(E_{q}+m_N)^2}\right)\frac{S_{12}({\bf q})}{6{\bf
q}^2}F^{(22)}(q',q;P),\tag{\theequation d} \label{eq56}
\end{align}
where
\begin{equation}S_{12}({\bf q})=3(\vec{\sigma}_1{\bf q})(\vec{\sigma}_2{\bf q})-{\bf q}^2(\vec{\sigma}_1\vec{\sigma}_2)\end{equation}
stands for the tensor operator and functions $F^{(ij)}(q',q;P)$ are defined as follows
\begin{align}
\stepcounter{equation}
&F^{(00)}(q',q;P)=\frac{1}{3}F^{(00)}_{\mu\nu}(q',q;P)\mathcal{P}^1_{\mu\nu},\tag{\theequation a}\\
&F^{(02)}=\frac{3}{\sqrt{2}}\widetilde{\Phi}_{(2\mathrm{s})}(q',P)\widetilde{\Phi}_{(2\mathrm{d})}(q,P)
\left[\widetilde{\mathcal{G}}_0(P^2)-\widetilde{\mathcal{G}}_2(P^2)\right]\sin\chi\cos\chi,\tag{\theequation b}\\
&F^{(20)}=\frac{3}{\sqrt{2}}\widetilde{\Phi}_{(2\mathrm{d})}(q',P)\widetilde{\Phi}_{(2\mathrm{s})}(q,P)
\left[\widetilde{\mathcal{G}}_0(P^2)-\widetilde{\mathcal{G}}_2(P^2)\right]\sin\chi\cos\chi,\tag{\theequation c}\\
&F^{(22)}=\frac{9}{2}\widetilde{\Phi}_{(2\mathrm{d})}(q',P)\widetilde{\Phi}_{(2\mathrm{d})}(q,P)
\left(\widetilde{\mathcal{G}}_0(P^2)\sin^2\chi+\widetilde{\mathcal{G}}_2(P^2)\cos^2\chi\right),\tag{\theequation
d}
\end{align}
and
\begin{equation}\label{eq85}\widetilde{\Phi}_{(2\mathrm{d})}(q,P)=\frac{2\beta}{\sqrt{15}(\beta+\alpha/2)^2}{\bf q}^2
\widetilde{\Phi}_{(0\mathrm{s})}(q,P).\end{equation}

It should be noted here that superscripts in the potentials $V^{(i,j)}$ {\em refer to the
dibaryonic channels}, i.e. they denote the angular quantum numbers of intermediate dibaryon.
However, as is clear from Eqs. (\ref{eq53})--(\ref{eq56}), due to the relativistic tensor
mixing, the angular momentum of the two-nucleon system is not conserved. This means that the
initial (or final) two-nucleon angular momentum (which is marked in the phase shift) may not
coincide with the angular momentum of the dibaryon. Therefore partial nucleon-nucleon potential
should be a sum of partial potentials corresponding to four different dibaryonic channels
(\ref{eq53})--(\ref{eq56}):
\[V^{110}_{L'L}=V^{(00)}_{L'L}+V^{(02)}_{L'L}+V^{(20)}_{L'L}+V^{(22)}_{L'L}.\]
Similar, albeit quite lengthy purely algebraic formulas have been derived for relativistic
$NN$-potentials in all other channels. It has been shown \cite{Ref20} that when doing the
nonrelativistic reduction of the above relativistic potentials we get the formulas very similar
to those obtained in our previous fully microscopic semi-relativistic approach
\cite{Ref6,Ref6a}.

After the complete derivation of the covariant $NN$-potential in various partial waves we should
add this short-range interaction to the peripheral one-pion and two-pion exchange potentials
derived previously in ChPT (but with the short-range contact terms being parameterized in a
purely phenomenological manner) or alternatively, we should replace the contact terms in the
ChPT-approach with the covariant short-range potential derived in our field-theoretical
approach. The resulting full $NN$-potential can be fitted first in the low-energy region,
$E_{\mathrm{lab}}\le 300$ MeV and by this way the main input parameters of the short-range part
of interaction (e.g. the coupling constants of the meson fields with the string) can be
established. This procedure is similar in some degree to fitting the parameters of contact terms
in higher-order ChPT. Thus, by combining the low-energy and high-energy effective field theories
one can reach a very consistent and fully dynamical description of elastic and inelastic
$NN$-interaction from zero energy up to the GeV region. This way is quite analogous to the
treatment of nucleon(or any other projectile)-nucleus scattering: at low energies and in lowest
partial waves the $s$-channel (or compound-state) scattering is dominating while at higher
partial waves the $t$-channel mechanisms should be prevailing. So, for a complete description of
this interaction we should combine somehow these two aspects.

\subsection{Illustrative model for $NN$ scattering in low partial waves}
We can now illustrate the applicability of the above EFT-approach by treating the lowest partial
$NN$ phase shifts with a simplified model derived from the general formulas
(\ref{eq21})-(\ref{eq85}) after the nonrelativistic reduction. For this purpose we will consider
the description for the $^1S_0$ and $^3S_1$-$^3D_1$ partial channels in the energy range from
zero up to 1 GeV. The total potential takes the form:
\begin{equation}
\label{eq86} V_{NN}^{\rm{total}}=V_{\rm{OPE}}+V_{\rm{TPE}}+V_{NqN},
\end{equation}
where $V_{\rm{OPE}}$ and $V_{\rm{TPE}}$ are the $t$-channel one- and two-pion exchange
potentials which are responsible for the peripheral part of $NN$-interaction. The OPE-potential
is taken here with soft cutoff constant $\Lambda_{\pi NN}=700$ MeV \cite{Ref6,Ref6a} which is
dictated by all dynamical models of $\pi N$-interaction. The $V_{\rm{TPE}}$ is approximated
here, for the sake of simplicity, by a simple local gaussian function
\[V_{\rm{TPE}}(r)=V_{\rm{TPE}}^{(0)}(\alpha r^2)^2\exp(-\alpha r^2).\]
The term $V_{NqN}$ in Eq.(\ref{eq86}) describes the intermediate- and short-range interaction
($r_{NN}\leq 1.2$ fm) induced by the dressed dibaryon formation in intermediate state.

\subsubsection{Description for $^1S_0$-channel}
After the nonrelativistic reduction of the general expression (\ref{eq21}), the $V_{NqN}$ part
of interaction in the uncoupled $^1S_0$ channel can be rewritten in the form:
\begin{equation}
\label{eq87} V_{NqN}({\bf q},{\bf q}')=\lambda^{(0)}f_{00}({\bf q})f_{00}({\bf
q}')+\lambda^{(2)}_{00}(E)f_{20}({\bf q})f_{20}({\bf q}'),
\end{equation}
where $f_{00}({\bf q})\equiv\langle{\bf q}|0\rm{s}\rangle$, $f_{20}({\bf q})\equiv\langle{\bf
q}|2\rm{s}\rangle$ are nonrelativistic h.o. functions of 0s and 2s form, $\lambda^{(0)}$ is an
arbitrary, but a large positive constant ($\sim 10^5-10^6$ MeV)\footnote{The first separable
term in Eq. (\ref{eq87}), according to our general approach, serves to represent the strong $NN$
repulsion when the intermediate six-quark bag has $|{\rm s}^6[6]_x\rangle$ structure
\cite{st,Stancu}.}. The energy-dependent strength parameter $\lambda^{(2)}_{00}(E)$ is
calculated straightforwardly from the $\sigma$-meson loop shown in Fig. \ref{fig4}. It was shown
earlier, the above loop integral can be well approximated in the energy range 0--1 GeV by a
simple Pade-approximant:
\begin{equation}
\lambda^{(2)}_{00}(E)=\lambda^{(2)}_{00}(0)\frac{E_0+aE}{E_0-E},
\end{equation}
with parameter values, $E_0$ and $a$, presented in Refs. \cite{Ref6a,Ref17}, while the strength
constant, $\lambda^{(2)}_{00}(0)$, is proportional to the $\sigma DD$ coupling constant which
determines the coupling of the $\sigma$-field and six-quark core. This constant is taken in this
illustrative model as a free parameter. So we have in the model only four free parameters, two
of which, $V_{\rm{TPE}}^{(0)}$ and $\alpha$, are related to the two-pion exchange potential in
the peripheral region, and the other two, $\lambda^{(2)}_{00}(0)$ and $r_0$ are related to the
dressed dibaryon properties. Two parameters of $V_{\rm{TPE}}$ are fixed easily by fitting two
effective range parameters which describe the $^1S_0$ phase shifts until 15 MeV. So, we are left
with only two basic parameters related to the properties of the $6q$-bag to describe the $^1S_0$
phase shifts in a very large energy range 15--1000 MeV. By varying these two parameters we
arrive at the high quality fit displayed in the Fig. \ref{fig9} (upper left panel).

\subsubsection{Description of $^3S_1$-$^3D_1$ coupled channels}
For the $^3S_1$-$^3D_1$ coupled channels we can derive the model from the general formulas by a
manner similar to the uncoupled $^1S_0$ case. Now, the dibaryon part of interaction, $V_{NqN}$,
can be presented in the matrix form:
\begin{figure}[t]
\begin{center}
\includegraphics[width=\textwidth,keepaspectratio]{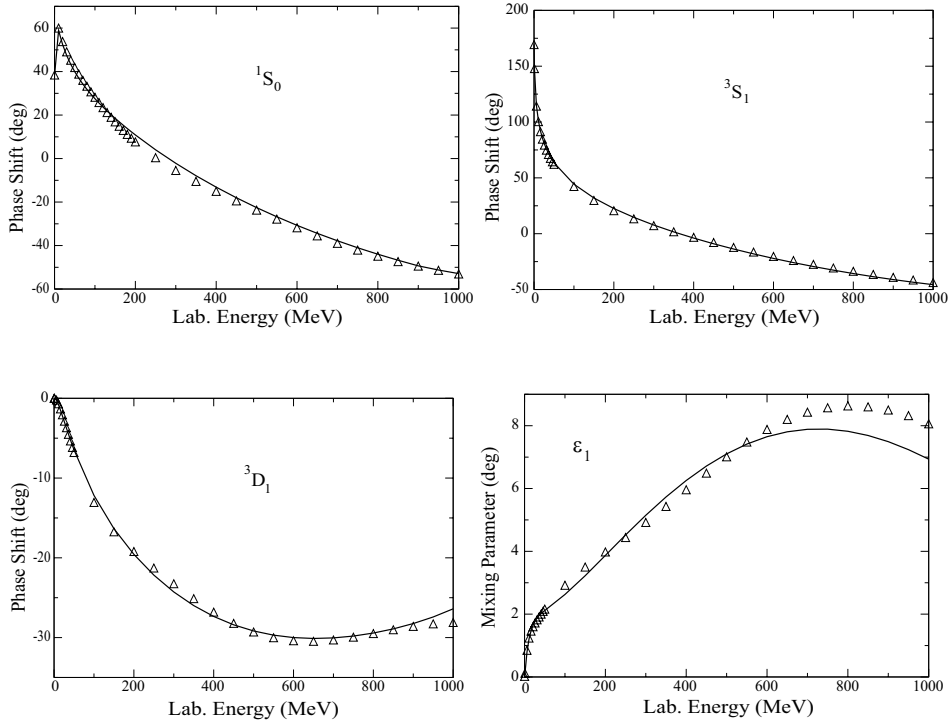}
\end{center}
 \caption{The comparison between the theoretical fits (with the $NN$
potential model similar to that given in eq. (\ref{eq21})) for some lower $NN$ phase shifts and
respective phase shift analysis (taken from SAID).\label{fig9}}
\end{figure}
\begin{equation}
\label{eq89} V_{NqN}({\bf q},{\bf q}')=\lambda^{(0)}f_{00}({\bf q})f_{00}({\bf
q}')+\widehat{V}_{NqN}^{(2\rm{s})}({\bf q},{\bf q}',E),
\end{equation}
where the matrix operator $\widehat{V}_{NqN}^{(2\rm{s})}$ takes the form:
\begin{equation}
\label{eq90} \widehat{V}_{NqN}^{(2\rm{s})}=\left(
\begin{array}{cc}
\lambda_{(00)}^{(2)}(E)f_{0}({\bf q})f_{0}({\bf q}')\ \ \ &\lambda_{(02)}^{(2)}(E)f_{0}({\bf
q})f_{2}({\bf
q}')\\
\lambda_{(20)}^{(2)}(E)f_{2}({\bf q})f_{0}({\bf q}')\ \ \ &\lambda_{(22)}^{(2)}(E)f_{2}({\bf
q})f_{2}({\bf q}')
\end{array}\right).
\end{equation}

One can attain a quite good description of the triplet phase shift in the whole energy range
0--1 GeV even when choosing the one common value of the oscillator radius for the $s$-wave,
$f_0({\bf q})$, and $d$-wave, $f_2({\bf q})$, potential form factors in Eq. (\ref{eq90}). The
parameter $a$, that determines the energy variation of all the strength parameters
$\lambda_{(ij)}^{(2)}$ is kept the same as in $^1S_0$-channel. So that in the $^3S_1$-$^3D_1$
coupled channel case, one has (except of two parameters in $V_{\rm{TPE}}$) only three adjustable
strength parameters, viz. $\lambda_{(00)}^{(2)},\ \lambda_{(02)}^{(2)}=\lambda_{(20)}^{(2)}$ and
$\lambda_{(22)}^{(2)}$ which make it possible to fit with a good quality the phase shifts
together with the respective tensor mixing parameter in a very large energy range 15--1000 MeV
(see Fig. \ref{fig9}, upper right and two lower panels)\footnote{If, in addition to these three
strength parameters, the $d$-wave oscillator parameter can be allowed to vary, the quality of
the fit may become almost perfect.}.

Thus, the illustrative model derived from EFT-approach developed in the work makes it possible,
using physically transparent interaction model with a minimal number of free parameters, to fit
very reasonably the phase shifts in $^1S_0$ and $^3S_1$-$^3D_1$ channels in the large energy
range and the deuteron structure as well. This description can be compared to the fit quality
for the phase shifts given by a purely phenomenological separable interaction (of conventional
type) of Graz group \cite{Graz}. The latter model includes more than twenty five free parameters
to fit the phase shifts in the same channels until 500 MeV only. Thus the description of the
elastic $NN$ scattering given by the present illustrative model derived from EFT-formalism looks
highly superior the description attained with the phenomenological separable models.

 It is worth to emphasize here that, at the value of the string excitation quantum
$\Omega\approx 300-350$ MeV, taking up to two-quanta excitations of the string will be
sufficient for description of $NN$-interaction from zero energy until 1 GeV or even higher,
while account of four-quanta excitations is sufficient for description of interaction in
$NN$-channel or the $NN$ meson-production processes up to energies of $E_{\mathrm{lab}}\approx
3$ GeV.

\section{Applications to other hadronic processes}
\subsection{Description of inelastic processes with the effective field theory approach}
Having developed the field-theoretical formalism for the description of the $NN$ elastic scattering in
the GeV region one can apply this also to describe inelastic processes
within the same framework, i.e. with the same input constants etc. And this possibility is one of main
advantage of this unified approach. For the sake of brevity we will omit many
details here and illustrate the description of inelastic processes by some graphs only.

In the above field-theoretical formalism the inelastic processes are tightly interrelated to the
elastic ones above the respective thresholds through the unitarity relation:
\begin{equation}T-T^\dag=2\mathrm{i}\,\mathrm{Im}TT^\dag,\end{equation}
which corresponds in our language to the respective cuts of the diagrams across the meson loops.

\begin{figure}[t]
\includegraphics[width=60mm,keepaspectratio]{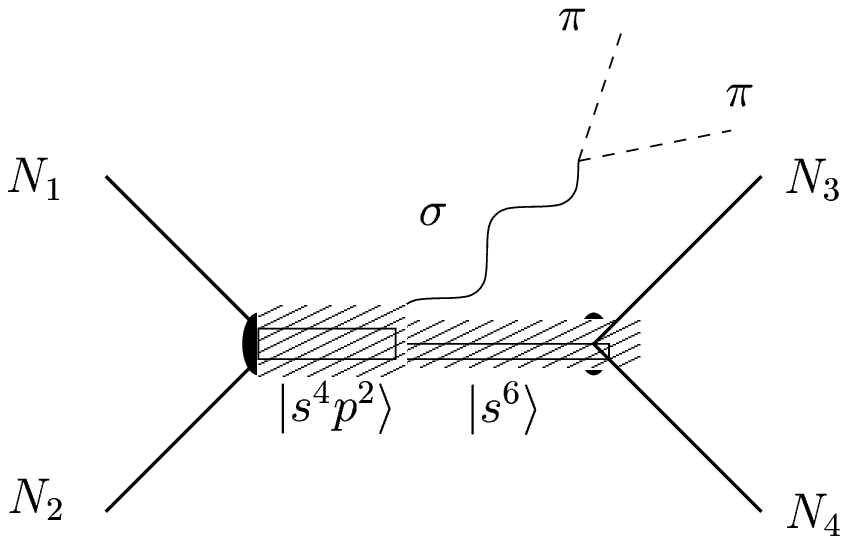} \hspace{1cm}
\includegraphics[width=60mm,keepaspectratio]{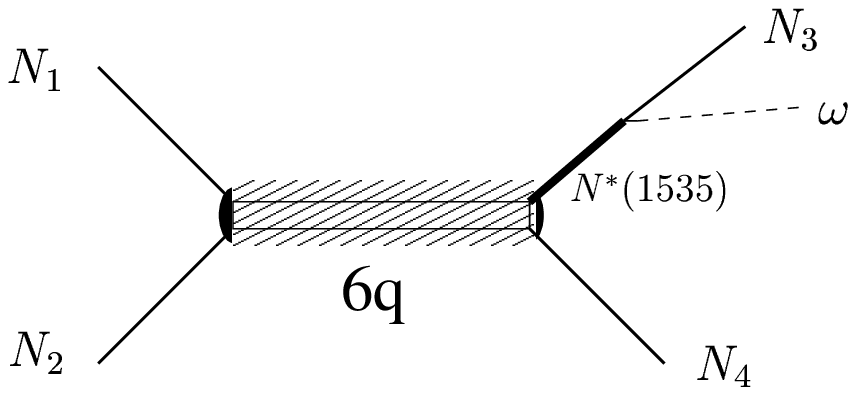}
\caption{Example of diagrams describing a new possible mechanisms of meson production. Left
diagram corresponds to double pion production in the scalar-isoscalar channel, whereas the right
one describes an enhancement of the probability for the formation of nucleon excited states and,
in particular, of the mechanism of $\omega$-meson emission in a nucleon-nucleon collision.
\label{fig10}}
\end{figure}

So, the respective meson production amplitudes can be illustrated by the graphs shown in Fig.
\ref{fig10}. The left graph relates to $2\pi$-production in $pp$-collisions through the
intermediate $\sigma$-meson in the dressed bag, while the graph on the right relates to the
near-threshold $\omega$-meson production to the intermediate $N^*$-isobar  generation. It should
be stressed here that the $NN$ ``distorted waves'' in such meson production amplitudes are
generated {\em from the same diagrams} (with the same parameter values) as these meson
productions. So the approach can be made quite consistent and universal both for elastic and
inelastic processes.

Another very important issue which can be answered on the basis of the developed approach is
multi-meson production and generation of the so called cumulative pions in high-energy $pp$
collisions. According to the relativistic picture presented here, the multi-meson production
should correspond to $N\hbar\Omega$ excitation of the intermediate string ($N>>1$) with its
subsequent deexcitation by multi-meson emission or by hadronic jets.

By a similar mechanism one can get the emission of cumulative pions (i.e. forbidden in a
single-nucleon emission), e.g. in the process $d+\mathrm{A}\rightarrow \pi+\ldots$, where the
energy of outgoing pion is larger then half energy of the incident deuterons.

\subsection{Treatment of electromagnetic processes at high momentum transfer and new electromagnetic
exchange currents}

The appearance of the dibaryon component in a nucleus can modify noticeably the nucleon momentum
distribution probed at large momentum transfer. E.g. one gets some new contributions to
electromagnetically induced two-nucleon knock-out processes in which the emerging two nucleons
(from the excited dibaryon decay) escape from the dibaryon state in ``back-to-back'' kinematics.
In fact, just this kinematics was observed for the two-nucleon ejection in many experiments of
this type. It is evident that such contributions for the $\mathrm{A} (e,e'pp)$ and
$\mathrm{A}(\gamma,pp)$ processes at high (virtual) momenta or energies of $\gamma$-quantum must
contribute to the reaction cross sections. In such two-nucleon processes, when treated within
our approach, the virtual high-energy (or momentum) can be absorbed by the string (i.e. by the
dibaryon as a whole) resulting in its excitation followed by either meson emission or break up
of the excited dibaryon into the two high-energy nucleons (see e.g. the Fig.~12). Also, the
inverse process of bremsstrahlung in $pp$-scattering: $pp\rightarrow pp\gamma$ should be
analysed with the same new currents.

\begin{figure}[t]
\begin{center}
\includegraphics{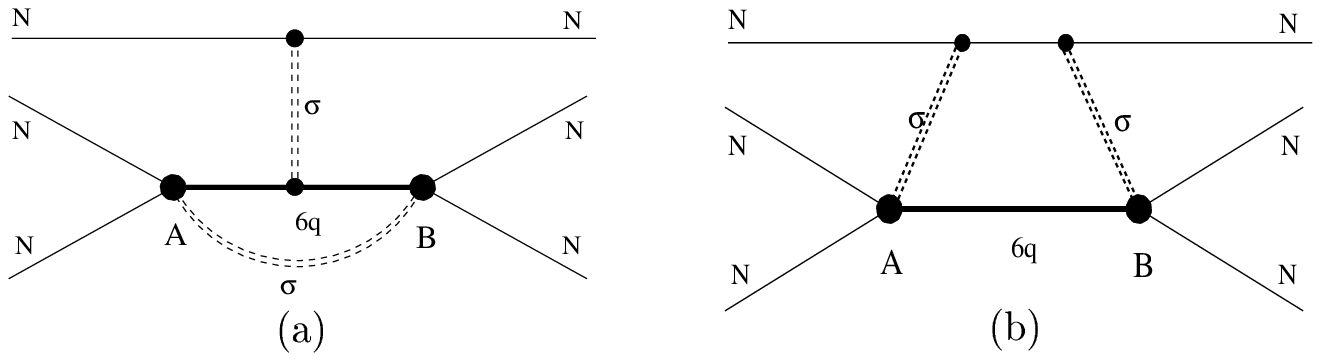}
\end{center}
 \caption{The graphs illustrating the new $3N$ scalar force induced by
$\sigma$-exchange between the dressed bag and third nucleon.\label{fig11}} 
\includegraphics[width=130mm,keepaspectratio]{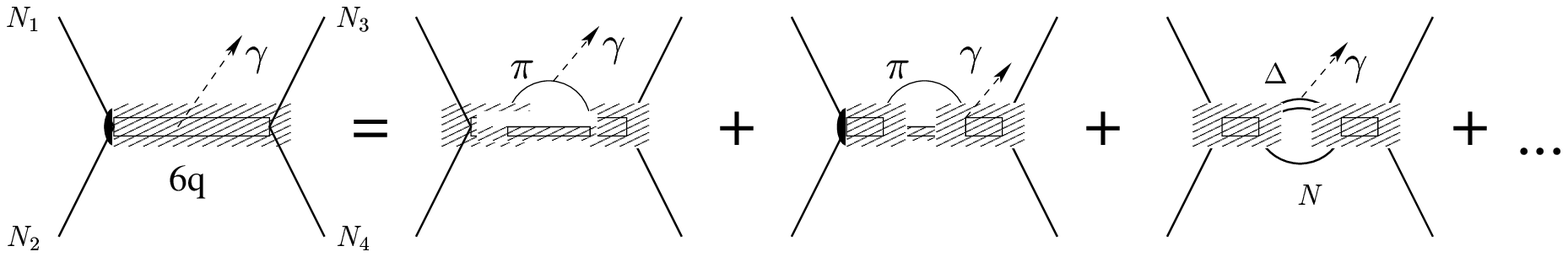}
\caption{Examples of
diagrams involving new meson-exchange electromagnetic currents that arise in the present model.
\label{fig12}}
\end{figure}

Another good test of the force model developed here will be in prediction of the main
characteristic features of e.-m. processes like $\nuc{12}{C}(\vec{\gamma},pN)$,
$\nuc{3}{He}(e,e'pN)$, or $\nuc{3}{He}(\vec{\gamma},pn)$ with linearly or circularly polarized
$\gamma$-quanta which are planned to be measured in Mainz \cite{Ref22}. For calculations of such
processes our very accurate $\nuc{3}{He}$ wavefunctions \cite{Ref17,Ref18} found with the above
force model can be employed jointly with new e.-m. currents tightly interrelated to the dressed
dibaryon degree of freedom. Some diagrams for these new $s$-channel currents in two-nucleon
systems are displayed in Fig. \ref{fig12}.

\begin{figure}[t]
\includegraphics[width=130mm,keepaspectratio]{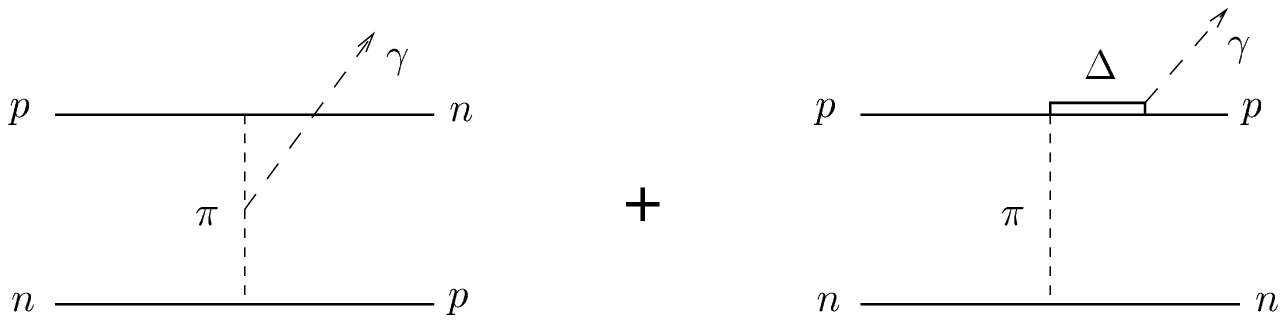}
 \caption{Example of
diagrams for $pn$ bremsstrahlung within traditional OBE-models of the nucleon-nucleon
interaction. \label{fig13}}
\end{figure}

The $s$-channel meson-exchange currents for $(\gamma,pp)$ or $(e,e'pp)$ processes should be
confronted with the traditional $t$-channel currents for $(pn)$-subsystem (see the Fig.
\ref{fig13}) and we can conclude from this comparison that both types of currents, i.e. $t$- and
$s$-channel ones can be considered as a complementary to each other. In other words, those
$t$-channel currents which disappear in the $pp$-system (i.e. pion-in-flight etc.)  will appear
in form of $s$-channel currents. The same occurs also for $pn$-system.

Thus the processes like $\nuc{3}{He}(e,e'pp)$ or $\nuc{3}{He}(e,e'pn)$ or
$\nuc{3}{He}(\gamma,pp)$ should be a very good test for the new electromagnetic currents. Hence,
the dressed dibaryon component and new currents associated with it should modify the results of
the theoretical interpretations of many dedicated experiments done at moderate to high momentum
and energy transfers, especially those which are now going on or are planned at Mainz and Bonn.

\subsection{Description of short-range $2N$ and $3N$ correlations in nuclei}
The first attempts to apply the DBS-model for description of properties of three-nucleon systems
\cite{Ref17,Ref18} was very successful. In particular, it has been clearly demonstrated that the
new short-range three-body attractive force induced by $\sigma$-meson exchange between the DBS
and third nucleon (see the Fig. \ref{fig11}) with the $g_{\sigma NN}$ coupling constant adjusted
to give the  $\nuc{3}{H}$ binding energy gives an important contribution to the $3N$ binding
energy and other properties of $3N$-system, and can explain quantitatively all static properties
of $\nuc{3}{He}$ and $\nuc{3}{H}$ ground states, including a precise parameter-free description
of the Coulomb displacement energy of $\nuc{3}{He}$--$\nuc{3}{H}$ \cite{Ref18}, and all the
charge distributions in these nuclei.

In these  $3N$ calculations done within the above dibaryon $NN$-force model with additional
incorporation of three-nucleon scalar forces shown by the graphs in the Fig.~\ref{fig11}, the
weight of the dibaryon component in the $3N$ wave function reaches as much as $10-11$\% and its
contribution to the total $3N$ binding energy is about 50\% (!). In other words, half of the
total nuclear binding comes from the strong interaction of dibaryon with surrounding nucleons
via the $\sigma$-meson exchange~\cite{Ref17}. Another important conclusion of these $3N$
calculations is a strong density dependence of the effective two- and three-body forces induced
by the intermediate dibaryon generation. The density dependence has a repulsive character and is
qualitatively similar to that assumed in the phenomenological Skyrme model. It is very likely
that just the interplay between the above density dependent repulsion and $2N$ and $3N$
attraction related to the scalar meson exchange provides the saturation property of nuclear
matter.

\section{Conclusions}
We developed in the work some covariant effective field-theoretical approach toward the
description of intermediate- and short-range components of $NN$ interaction and nuclear forces.
The approach assumes that the nucleon-nucleon interaction at intermediate and short ranges has a
very complicated and multistep character. In the first step, the two-nucleon system goes, by
color exchange, to the confined two-cluster $CC$ configuration where the quark clusters are
moving at the end of a quantum string (inside of which the gluon fields are localized).
Excitations (i.e. vibrations and rotations) of this string can be described by a relativistic
covariant harmonic oscillator with a linear confinement potential which leads quite naturally to
the linear Regge-trajectory of the respective excited levels of the string. In the next step,
the string interacts with vacuum meson fields (mainly $\pi,\sigma$ and $\omega$) and thus it
goes to the dibaryon stage by the respective dressing. It is shown that
$2\hbar\Omega$-excitation of the string generates a strong scalar $\sigma$-field which, due to
the strong attraction to the six-quark core, will compress and stabilize the resulting dressed
six-quark bag.

In some aspect the emerged picture should be rather similar to the concept of ``small chiral
bag'' suggested long ago by Brown and Rho \cite{Brown}. In their model the $3q$-MIT bag is
compressed by the strong pressure of its pion (chiral) field to the small size where the
additional inner kinetic pressure of free quarks will compensate the external pressure of the
pion field to the bag surface. In our case, in contrast to the above ``small bag model'', the
main meson field is a strong scalar $\sigma$-field which interacts with the six quark system in
the whole volume of the bag rather than on the its surface only. Another great difference from
the Brown-Rho model is the origin of the stabilizing field ($\sigma$ in our case). In our
approach the main component of this scalar field is emerged in deexcitation of the string from
$2\hbar\Omega$ to $0\hbar\Omega$ states. Thus, in this process one observes the transformation
of the ``color'' energy of gluon fields (inside the string) to the ``white'' energy of scalar
meson field.

In the next stage of the process, the $\sigma$ field will compress and stabilize the six-quark
bag and shift strongly its mass down to the pion-production threshold. Thus this shift should be
included to the polarization operator $\Pi(P^2)$ of the total string propagator. Further,
keeping in mind a strong coupling of the $\sigma$-meson to the two-pion channel one can predict
from this specific dressing mechanism an enhanced two-pion production in $NN$ collision as
compared to the traditional OBE-models, the feature just observed experimentally. Moreover,
almost the same mechanism can enhance quite naturally the single-pion production in
$pp$-collision and especially the near threshold $\pi_0$ production. The consistent theoretical
explanation of which met so serious problems in the recent decade \cite{Ref9}. Moreover, a quite
similar two-stage mechanism (in first stage, the excitation of the color string with the
subsequent transformation of the ``color'' string energy to the meson fields) should be
responsible for multi-meson production in $NN$ and heavy ion collisions. So, by this way, the
multi-meson production in high-energy heavy-ion collision may give a good test for the proposed
color string model. The authors plan to return to this point in their next papers.
\begin{ack}
The authors appreciate very much the fruitful discussions in the course of this work with P.
Grabmayr, M. Kaskulov, I.T. Obukhovsky and V.N. Pomerantsev. The work of the V.I.K. and M.A.S.
is partially supported financially by the DFG-grant no. Fa-67/20-1 and RFBR-grants nos.
02-02-16612 and 04-02-04001.
\end{ack}


\newpage
\begin{thebibliography}{99}
\bibitem{Afnan} A.D. Lahiff, I.R. Afnan, Phys. Rev. C 60 (1999) 024608.
\bibitem{ref1} A. Nogga, H. Kamada, W. Gl\"ockle, Phys. Rev. Lett. 85 (2000) 944.
\bibitem{ref2} V.I. Kukulin, I.T. Obukhovsky, V.N. Pomerantsev, A. Faessler, Phys. At. Nucl. 64 (2001) 1667.
\bibitem{ref3} D. Plaemper, J. Flender, M.F. Gari, Phys. Rev. C 49 (1994) 2370.
\bibitem{ref3a} T. Sasakawa, S. Ishikawa, Few Body Syst. 1 (1986) 3.
\bibitem{Ref4} R.S. Hicks et al., Phys. Rev. C 67 (2003) 064004;
F. Moschini, Proc. of 6th Workshop on ``e-m induced Two-Hadron Emission'', Pavia, September
24-27, 2003, p. 156.
\bibitem{Groep} D.L. Groep et al., Phys. Rev. C 63 (2001) 014005.
\bibitem{Ref13} S. Ishikawa, J. Golak, H. Witala, H. Kamada, W. Gl\"ockle, D. H\"uber, Phys. Rev. C 57 (1998) 39;
R. Skibi\'nski et al., Phys. Rev. C 67 (2003) 054002; W. Gl\"ockle et al., Proc. of 6th Workshop
on ``e-m induced Two-Hadron Emission'', Pavia, September 24-27, 2003, p. 166.
\bibitem{Ref5} E. Jans, P. Barneo,  Proc. of 6th Workshop on ``e-m induced Two-Hadron Emission'',
Pavia, September 24-27, 2003, p. 183.
\bibitem{Ref5a} L.B. Weinstein, Proc. of 5th Workshop on ``e-m induced Two-Hadron Emission'', Lund, June 13-16, 2001, p. 93;
E. Piasetzky, R. Gilman, M. Sargsian, Proc. of 6th Workshop on ``e-m induced Two-Hadron
Emission'', Pavia, September 24-27, 2003, p. 211.
\bibitem{Laget} J.-M. Laget, Nucl. Phys. A 579 (1994) 333.
\bibitem{st} F. Stancu, Few Body Syst. Suppl. 14 (2003) 83; D. Bartz, F. Stancu, Phys. Rev. C 63 (2001) 034001;
Nucl. Phys. A 699 (2002) 316.
\bibitem{VM1} R.V. Mau, Prog. Part. Nucl. Phys. 50 (2003) 561.
\bibitem{Yaz} M. Oka, K. Shimizu, K. Yazaki, Prog. Theor. Phys. Suppl. 137 (2000) 1.
\bibitem{Sal}  D.R. Entem, F. Fern\'andez, A. Valcarce, Phys. Rev. C 62 (2000) 034002; C 67 (2003) 014001.
\bibitem{Weinberg} S. Weinberg, Phys. Lett. B 251 (1990) 288; Nucl. Phys. B 363 (1991) 3.
\bibitem{Ref7} D.B. Kaplan, nucl-th/9506035; Nucl. Phys. B 494 (1997) 471; D.B. Kaplan, M. Savage, M.B. Wise,
Nucl. Phys. B 478 (1996) 629; Phys. Lett. B 424 (1998) 390.
\bibitem{EntMach} D.R. Entem, R. Machleidt, Phys. Rev. C 68 (2003) 041001.
\bibitem{Epel} E. Epelbaum, W. Gl\"ockle, U.G. Meissner, Eur. Phys. J. A 19 (2004) 125.
\bibitem{Higa} R. Higa, M.R. Robilotta, Phys. Rev. C 68 (2003) 024004;
R. Higa, M.R. Robilotta, C.A. da Rocha, Phys. Rev. C 69 (2004) 034009.
\bibitem{Ref8} H.O. Meyer et al.,  Phys. Rev. C 63 (2001) 064002.
\bibitem{Ref9} C. Hanhart, Phys. Rep. 397 (2004) 155.
\bibitem{Ref9a} B. von Przewoski et al., Phys. Rev. C 58 (1998) 1897.
\bibitem{Ref9b} W.W. Daehnick et al., Phys. Rev. C 65 (2002) 024003.
\bibitem{Ref10} Bo H\"oistad, Nucl. Phys. A 721,  (2003) 570c;
J. Johanson et al., Nucl. Phys. A 712 (2002) 75.
\bibitem{Shimizu}F. Shimizu et. al., Nucl. Phys. A 386 (1982) 571.
\bibitem{Ref10a} L. Alvarez-Ruso, E. Oset, E. Hernandez, Nucl. Phys. A 633 (1998) 519.
\bibitem{Brod} S.J. Brodsky, L. Frankfurt, R. Gilman, J.R. Hiller, G.A. Miller, E. Piasetzky, M. Sargsian,
M. Strikman, Phys. Lett. B 578 (2004) 69.
\bibitem{Ref6} V.I. Kukulin, V.N. Pomerantsev, A. Faessler, J. Phys. G 27 (2001) 1851.
\bibitem{Ref6a} V.I. Kukulin, I.T. Obukhovsky, V.N. Pomerantsev, A. Faessler, Int. J. Mod. Phys. E 11 (2002) 1.
\bibitem{Ny} E.M. Nyman, D.O. Riska, Phys. Scripta 34 (1986) 533; Phys. Lett. B 203 (1988) 13.
\bibitem{Oset} E. Oset, H. Toki, M. Mizobe, T.T. Takahashi, Progr. Theor. Phys. 103 (2000) 351.
\bibitem{Kask} M.M. Kaskulov, H. Clement, Phys. Rev. C 70 (2004) 014002.
\bibitem{Kais}N. Kaiser, U.G. Meissner, Phys. Lett. B 233 (1989) 457.
\bibitem{Ref14} V.I. Kukulin, Few Body Syst. Suppl. 14 (2003) 71.
\bibitem{Ref17} V.I. Kukulin, V.N. Pomerantsev, M.M. Kaskulov, A. Faessler, J. Phys. G 30 (2004) 287.
\bibitem{Ref18} V.I. Kukulin, V.N. Pomerantsev, A. Faessler, J. Phys. G 30 (2004) 309.
\bibitem{Obukh} I.T. Obukhovsky, V.I. Kukulin, M.M. Kaskulov, P. Grabmayr, A. Faessler, J. Phys. G 29 (2003) 2207.
\bibitem{Ref20} V.I. Kukulin, M.A. Shikhalev, Phys. At. Nucl. 67 (2004) 1536.
\bibitem{Ref19} M.M. Kaskulov, P. Grabmayr, V.I. Kukulin, Int. J. Mod. Phys. E 12 (2003) 449;
M.M. Kaskulov, V.I. Kukulin, P. Grabmayr, Few Body Syst. Suppl. 14 (2003) 101.
\bibitem{Obukh2} A.M. Kusainov, V.G. Neudatchin, I.T. Obukhovsky, Phys. Rev. C 44 (1991) 2343.
\bibitem{Shuryak}P. Faccioli, E.V. Shuryak, Phys. Rev. D 65 (2002) 076002.
\bibitem{Rapp} R. Rapp, T. Schafer, E. Shuryak, M. Velkovsky, Phys. Rev. Lett. 81 (1998) 53.
\bibitem{Alford}M. Alford, K. Rajagopal, F. Wilczek, Phys. Lett. B 422 (1998) 247.
\bibitem{Huang}M. Huang, P.F. Zhuang, W.Q. Chao, Phys. Rev. D 65 (2002) 076012.
\bibitem{Wilczek} F. Wilczek, hep-ph/0409168.
\bibitem{Stancu}D. Bartz, Fl. Stancu, Phys. Rev. C 59 (1999) 1756.
\bibitem{Bjorken} J.D. Bj\"{o}rken, S.D. Drell, Relativistic Quantum Mechanics, McGraw--Hill,
New York, 1964.
\bibitem{Ishida} R.P. Feynman, M. Kislinger, F. Ravndal, Phys. Rev. D 3 (1971) 2706;
S. Ishida, M. Ishida, T. Maeda, Prog. Theor. Phys. 104 (2000) 785.
\bibitem{Ref21} M. Moshinsky, A. Szczepaniak, J. Phys. A 22 (1989) L817;
A. Szczepaniak, A.G. Williams, Phys. Rev. D 47 (1993) 1175.
\bibitem{Takabayasi}T. Takabayasi, Phys. Rev. B 139 (1965) B1381; Suppl. Prog. Theor. Phys. 67
(1979) 1.
\bibitem{Graz} L. Mathelitsch, W. Plessas, M. Schweiger, Phys. Rev. C 26 (1982) 65.
\bibitem{Ref22} C.J.Y. Powrie et al., Phys. Rev. C 64 (2001) 034602.
\bibitem{Brown} G.E. Brown, M. Rho, Comments Nucl. Part. Phys. 18 (1988) 1; see also
S.W. Hong and B.K. Jennings, Phys. Rev. C 64 (2001) 038203.
\end{thebibliography}
\end{document}